\documentclass{ieeetmlcn}
\usepackage{cite}
\usepackage{amsmath,amssymb,amsfonts}
\usepackage{algorithmic}
\usepackage{graphicx,color}
\usepackage{textcomp}
\usepackage{xcolor}
\usepackage[hidelinks]{hyperref}
\usepackage{algorithm,algorithmic}
\usepackage{multirow}
\usepackage[nowarn,acronyms,nonumberlist,nopostdot,nomain,nogroupskip]{glossaries}
\usepackage{subcaption}
\let\labelindent\relax
\usepackage[inline]{enumitem} 
\usepackage{tabularx}
\usepackage{comment}
\usepackage{soul}
\usepackage{makecell}
\graphicspath{{images/}}
\newacronym{ra}{RA}{random access}
\newacronym{rl}{RL}{reinforcement learning}
\newacronym{aop}{AoP}{age of packet}
\newacronym{aoi}{AoI}{age of information}
\newacronym{eb}{EB}{exponential backoff}
\newacronym{beb}{BEB}{binary exponential backoff}
\newacronym{dqn}{DQN}{deep Q-network}
\newacronym{drl}{DRL}{deep reinforcement learning}
\newacronym{drqn}{DRQN}{deep recurrent Q-network}
\newacronym{fs}{FS}{fixed scheduling}
\newacronym{ed}{ED}{event-driven}
\newacronym{htc}{HTC}{human-type communication}
\newacronym{mtc}{MTC}{machine-type communication}
\newacronym{mmtc}{mMTC}{massive machine-type communication}
\newacronym{mtd}{MTD}{machine-type device}
\newacronym{bs}{BS}{base station}
\newacronym{marl}{MARL}{multi-agent reinforcement learning}
\newacronym{maddpg}{MADDPG}{multi-agent deep deterministic policy gradient}
\newacronym{mappo}{MAPPO}{multi-agent proximal policy optimization}
\newacronym{coma}{COMA}{counterfactual multi-agent policy gradients}
\newacronym{mfmarl}{MF-MARL}{mean field MARL}
\newacronym{maac}{MAAC}{multi-actor-attention critic}
\newacronym{vdn}{VDN}{value decomposition networks}
\newacronym{qmix}{}{QMIX}
\newacronym{rnn}{RNN}{recurrent neural network}
\newacronym{prb}{PRB}{physical resource block}
\newacronym{gru}{GRU}{gated recurrent unit}
\newacronym{ps}{PS}{parameter sharing}
\newacronym{nbiot}{NB-IoT}{narrowband IoT}
\newacronym{ltem}{LTE-M}{LTE-Machine-type communication}
\newacronym{lpwan}{LPWAN}{low power wide area network}
\newacronym{ctde}{CTDE}{centralized training and decentralized execution}

\def\BibTeX{{\rm B\kern-.05em{\sc i\kern-.025em b}\kern-.08em
    T\kern-.1667em\lower.7ex\hbox{E}\kern-.125emX}}
\AtBeginDocument{\definecolor{tmlcncolor}{cmyk}{0.93,0.59,0.15,0.02}\definecolor{NavyBlue}{RGB}{0,86,125}}

\def\authorrefmark#1{\ensuremath{^{\textbf{#1}}}}
\newtheorem{definition}{Definition}

\begin{document}

\markboth{}{Muhammad Awais Jadoon {et al.}}

\title{Learning Random Access Schemes for Massive Machine-Type Communication with MARL}

\author{Muhammad Awais Jadoon\authorrefmark{1}, Student Member, IEEE, Adriano Pastore\authorrefmark{1}, Senior Member, IEEE,  Monica Navarro\authorrefmark{1}, Senior Member, IEEE, and Alvaro Valcarce\authorrefmark{2}, Senior Member, IEEE}
\affil{Centre Tecnològic de Telecomunicacions de Catalunya (CTTC)/CERCA, Castelldefels, Barcelona}
\affil{Department of Radio Systems Research \& AI, Nokia Bell-Labs France, Massy, France}

\corresp{Corresponding author: Muhammad Awais Jadoon (email: mjadoon@cttc.es).}

\begin{abstract}
In this paper, we explore various \gls{marl} techniques to design grant-free \gls{ra} schemes for low-complexity, low-power battery operated devices in \gls{mmtc} wireless networks. We use \gls{vdn} and QMIX algorithms with \gls{ps} with \gls{ctde} while maintaining scalability. We then compare the policies learned by \gls{vdn}, QMIX, and \gls{drqn} and explore the impact of including the agent identifiers in the observation vector. We show that the \gls{marl}-based \gls{ra} schemes can achieve a better throughput-fairness trade-off between agents without having to condition on the agent identifiers. We also present a novel correlated traffic model, which is more descriptive of \gls{mmtc} scenarios, and show that the proposed algorithm can easily adapt to traffic non-stationarities.
\end{abstract}

\begin{IEEEkeywords}
Massive machine-type communications, MARL, reinforcement learning, grant-free random access, scalability.
\end{IEEEkeywords}


\maketitle

\section{INTRODUCTION}
\IEEEPARstart{T}{he} \gls{mmtc} paradigm is a key component of 5G and will continue to be important in the development of 6G technologies \cite{five_5g}.  As the number of Internet of Things (IoT) devices grows, millions of devices with characteristics different from human-type communication will require connectivity \cite{bockelmann_18, bockelmann_16}. To support \gls{mmtc} in LTE-A, 3GPP has developed \gls{nbiot} and \gls{ltem} \cite{nbiot_ltem}, which fall under the category of \glspl{lpwan}. In addition to these cellular standards, non-cellular \gls{lpwan} standards such as Sigfox \cite{sigfox} and LoRa \cite{lora} have also been developed. 3GPP's Rel-17 introduces 'NR-Light', a new class of devices that is more capable than \gls{nbiot} or \gls{ltem}, but supports different features with a bandwidth larger than \gls{nbiot}/\gls{ltem} but smaller than 5G NR devices. In this paper, we focus on low-power, low-complexity \glspl{mtd} with low data rates (around 1-100 Kbps), where communication is mostly uplink dominated. These devices are also low-cost and battery-operated, with long battery life and sporadic activity. Managing medium access for these devices is challenging, and future wireless communication systems will need to provide massive connectivity to meet these needs.

For devices having such characteristics, grant-free \gls{ra} schemes are preferred as scheduled access incurs huge signaling overhead \cite{ra_mtc_israel_19}. However, \gls{ra} schemes are prone to collisions and scale poorly. Traditionally, \gls{ra} schemes such as \gls{eb} \cite{beb_analysis} employ back-off mechanism at each device to update their transmit probabilities based on the feedback from the receiver. These schemes are relatively simple and decentralized; however, their performance depends on various assumptions such as the traffic arrival process and whether the buffers are saturated. Additionally, the optimal back-off factor for different system parameters is not fixed and may vary \cite{Barletta18}. One drawback of \gls{eb} schemes such as \gls{beb} is the \emph{capture effect}, where a group of devices occupy the channel for a period of time, causing other devices to be deprived of access and making the technique unfair. Our goal is to design \gls{ra} schemes for \gls{mmtc} that not only provide better throughput but are also fair.

\Gls{rl} algorithms have become a popular method for learning \gls{ra} policies in wireless networks. These algorithms can adapt to changes in the environment and use past history to learn the transmission probabilities of devices in a decentralized manner. However, many \gls{rl} solutions are not tailored to the traffic and device characteristics of \gls{mmtc} systems (details in Section \ref{sec:marl_suitability}), and they also struggle with scalability to large numbers of devices, which is a critical concern in \gls{ra} schemes for \gls{mmtc}. This is because \gls{mmtc} devices often have low computational power and rely on battery power, making it impractical to perform learning at each device.  To the best of our knowledge, the scalability issue has not been adequately addressed in previous \gls{rl} or \gls{marl} studies on \gls{ra} schemes, and it is unclear if the proposed techniques can handle a large number of devices. \gls{marl} has several advantages over traditional \gls{eb} backoff policies, as it allows for the design of multi-objective policies in a decentralized manner, which is not analytically tractable using traditional methods.

Therefore, the objective of this paper is to design grant-free \gls{ra} schemes for \gls{mmtc} using \gls{marl} to achieve fairness, adaptability to changes in traffic, centralized learning with decentralized execution, and scalability to a large number of devices. Our contributions are listed in the following. 

\begin{itemize}
    \item We present a system model to learn schemes in which the devices can leave and join the network randomly. We do not assume that all the devices in the network have packets as opposed to most of the other works for \gls{ra} with \gls{rl}, e.g., \cite{Naparstek, wang2018}.
    \item We use broadcast feedback to reduce the signaling overhead and for energy efficiency. We assume that the feedback is only sent to the active devices and to save energy, the inactive devices do not listen to the feedback signal. 
    \item We present a suitability report of \gls{marl} algorithms for our proposed environment. Since we want a single policy for all the devices that can be learned in a \gls{ctde} manner; we provide a comparison between some well-known \gls{marl} algorithms and how they may or may not be suitable for our environment. We propose \gls{vdn} and QMIX algorithms to achieve our objectives. We present our simulation results for \gls{vdn} for a multiple-user multiple \glspl{prb} environment and also compare \gls{vdn}, QMIX and \gls{drqn} policies.
    \item Most of the \gls{marl} algorithms that employ \gls{ctde}, include an agent-specific identifier into the observation vector of the agents. In case of \gls{mmtc}, the devices should be able to leave/join the network and the policies should be scalable to a large number of devices. For these reasons, incorporating agent/device identification (ID)  is not feasible. We will also show that how the algorithm distributes resources among \glspl{mtd} fairly when agent IDs are not incorporated and how the algorithms learn an unfair policy when we use agent IDs.
    \item We present our results for \emph{regular} or periodic traffic arrival, in which each \gls{mtd} receives packets following a random process independently. In addition, we present a correlated traffic arrival model in Section \ref{sec:correlated_model}, that is more suitable for \gls{mmtc} system. In the correlated traffic arrival model, the devices follow both regular traffic arrivals and also \gls{ed} traffic arrival. In \gls{ed}, that is independent of the regular traffic arrival, a subset of \glspl{mtd} become active together whenever an event happens. We show that our proposed algorithm adapts to different traffic conditions.
\end{itemize} 

\section{RELATED WORK}
The application of \gls{rl} to channel access problems in wireless communications goes back to 2010 that used tabular Q-learning \cite{HLi}. However, it has become popular in the recent years due to the advancements in \gls{drl}. In \cite{challita_drl}, the authors considered the problem of multiple access where the agents are the base stations to predict the future state of the system. They use \gls{rnn} and REINFORCE algorithm to learn policies for each agent. In \cite{CHU201523}, the ALOHA-Q protocol is proposed for a single channel slotted ALOHA scheme that uses an expert-based approach in \gls{rl}. The goal in that work is for nodes to learn in which time slots the likelihood of packet collisions is reduced. However, the ALOHA-Q depends on the frame structure and each user keeps and updates a separate policy for each time slot in the frame. In \cite{alfaro_alohaq}, the ALOHA-Q is enhanced by removing the frame structure. However, every user still has to keep the number of policies equal to the time slots window it is going to transmit in. Other works such as \cite{Naparstek, zhong2019deep, wang2018, xu_milcom} consider \gls{rl}-based multiple access works for multiple channels. In \cite{Naparstek, tomovic_20, wang2018} \gls{dqn} algorithm is used for multiple user and multiple channel wireless networks. In \cite{zhong2019deep}, another \gls{drl} algorithm known as \textit{actor-critic} \gls{drl} is used for dynamic channel access. All of these works train agents with the assumption that every device has always a packet in its buffer (saturation state). Moreover, it is not clear whether their algorithms can be scaled for higher number of agents. Interestingly, these works also do not compare their results with any backoff techniques such as \gls{eb} to show whether their results outperform them. 
In \cite{rapid}, authors propose a \gls{ra} procedure for delay sensitive applications using the context IDs of the devices along with the two-step \gls{ra} procedure. This is done by predicting the traffic of the devices. A \gls{ra} strategy for initial access (4 message exchange) to allocate resources in proposed in \cite{nan_21}. They assume that each device also reports its energy levels and access delay to the centralized receiver. Therefore, signaling overhead in this work is high for massive access and it is not energy efficient. 

Recently, a \gls{ra} protocol for initial access is proposed in \cite{zhang_doubleRA22} where results were shown for both regular and bursty traffic arrivals. A \gls{rl}-based strategy has been proposed in \cite{alberto_21} for the correlated traffic model. In our previous work \cite{jadoon_wcnc22}, we had used \gls{dqn} with a single resource for the devices following Poisson process for traffic arrival and in \cite{jadoon_vtc} we showed how \gls{dqn} with \gls{ps} is scalable for bursty traffic arrival. To show the effectiveness of \gls{drl} in learning new access strategies, in \cite{8Yu2019het}, a heterogeneous environment is considered in which an \gls{rl} agent learns an access scheme in co-existence with slotted ALOHA and a time division multiple access (TDMA) access schemes. In \cite{acb_nbiot}, access class barring (ACB) mechanism has been optimized for \gls{nbiot} using \gls{drl}. A multiple access algorithm is designed using actor-critic \gls{marl} in \cite{Shao_22}. 
\begin{figure}
    \centering
    \includegraphics[width=200px, height=160px]{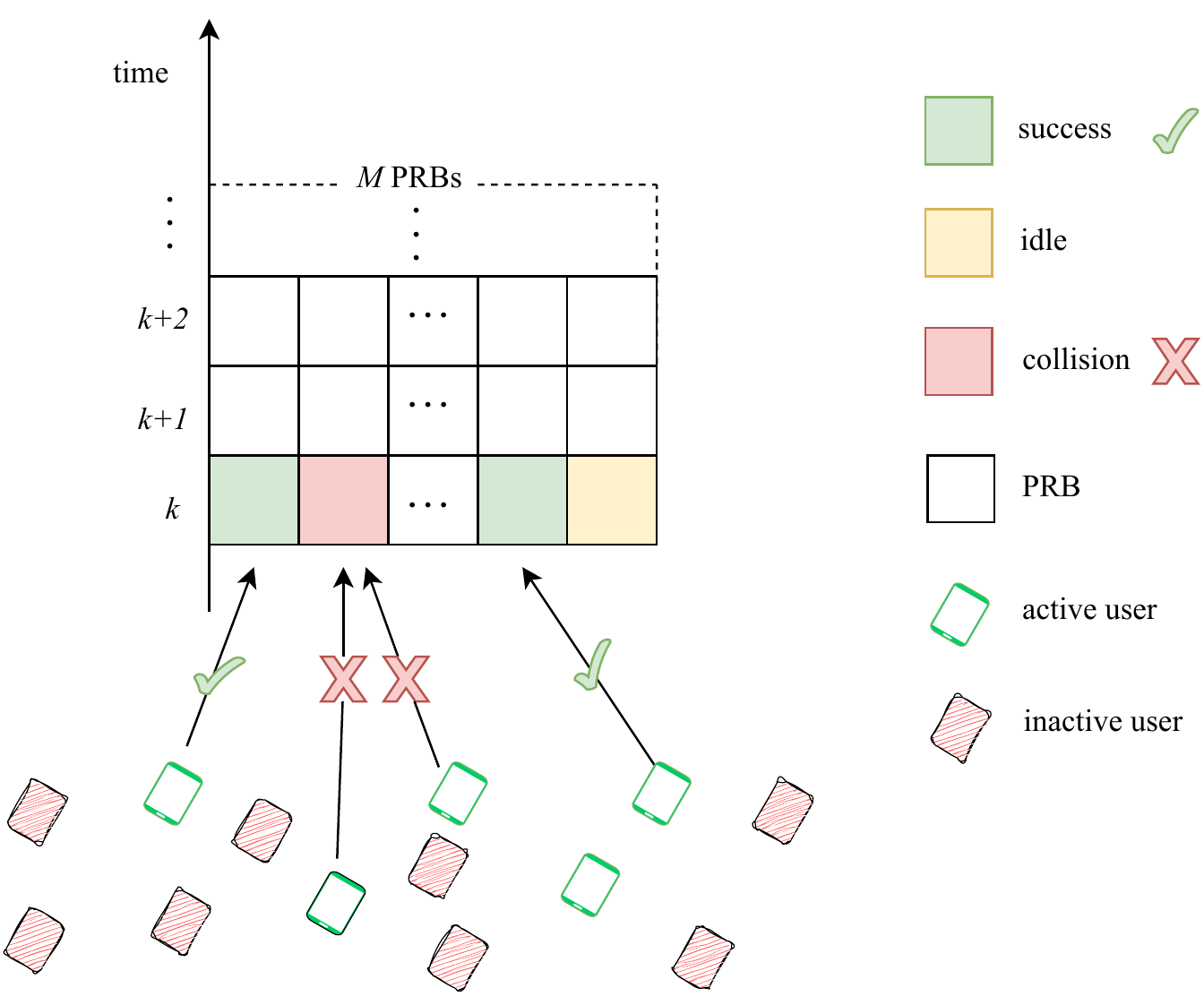}
    \caption{System Model}
    \label{fig:system_model_mtc}
\end{figure}

\section{SYSTEM MODEL AND PROBLEM FORMULATION}
We consider a synchronous time-slotted wireless network with a set $\mathcal{N} = \{1, \dotso, N \}$ of \glspl{mtd}, a set $\mathcal{M} = \{1, \dotso, M\}$ of shared orthogonal \glspl{prb} and a receiver as shown in Fig. \ref{fig:system_model_mtc}. The physical time is divided into slots, each of duration $1$ and the slot index is $k \in \mathbb{N}$. At each time slot, we assume that only $\mathcal{N}_a \subseteq \mathcal{N}$ devices are active and the activity pattern follows a random process. Each active \gls{mtd} transmits over the shared \glspl{prb} in a grant-free manner. At each time slot $k$, an \gls{mtd} can transmit only one packet and it can transmit it only over one resource $m \in \mathcal{M}$. The \glspl{mtd} are assumed to have perfect synchronization. 
Moreover, each \gls{mtd} is equipped with a buffer to store the packets in its queue  and each device $n$ can only store at most one packet. The buffer state at time $k$ is defined as $B_n(k) \in \{0,1\}$, where $B_n(k)=1$ if there is a packet in the buffer and it is $0$ otherwise. If the buffer $B_n(k)$ is full, new packets arriving at device $n$ are discarded and are considered lost. Each device becomes active whenever a packet is generated at the device following one of the traffic arrival models given in Section \ref{sec:correlated_model}.
At each time slot $k$, \gls{mtd} $n$ takes an action 
\begin{align}
   A_n(k) \in \mathcal{A} = \{0,1, \dotso, M\}, 
\end{align}
where $A_n(k)=0$ corresponds to the event when user $n$ chooses to not transmit and $A_n(k)=m$ corresponds to the event when user $n$ transmits a single packet on channel $m$ for $1 \leq m \leq M$. 
If only one user transmits on the channel $m$ in a given time slot $k$, the transmission is successful, whereas a collision event happens if two or more devices transmit in the same time slot. The collided packets are discarded and need to be retransmitted until they are successfully received at the receiver. 

For feedback, we consider a broadcast feedback signal $F(k)$ from the receiver that is common to all the devices. Formally, we define
\begin{align}
   F(k) = \{F_1(k), \dotso, F_M(k) \} ,
\end{align}
and $F_m(k)$ stands for the feedback corresponding to the channel $m$ and for each time slot $k$, it is defined as  
\begin{align}\label{eq:local_feedback}
        F_m(k) &= 
        \begin{cases}
    		1 & \text{if success at time slot $k$}\\
    		0 & \text{otherwise}\\
    	\end{cases} 
\end{align}

Let us define the binary set $\mathcal{B} = \{0, 1\}$. We define \emph{success} and \emph{collision} event for user $n$ as a function of the feedback $F_m(k)$ and $A_{n}(k)$, i.e., $g:(F_m(k), A_n(k)) \mapsto G_{n,m}(k)$, and $C_{n,m}(k)$, where $G_{n,m}(k)\in \mathcal{B}$ and $C_{n,m}(k) \in \mathcal{B}$ are the success and collision event for the device $n$ respectively, and they are locally computed by each device. Formally, we define the success event for user $n$ and $\forall m \in \mathcal{M}$ as, 

\begin{align} \label{eq:success_event}
        G_{n,m}(k) &= 
        \begin{cases}
    		1 & \text{if $A_n(k) = m$ and $F_m(k)=1$}\\
    		0 & \text{otherwise,}
    	\end{cases} 
    \end{align}
and the collision event as,
\begin{align} \label{eq:collision_event}
        C_{n,m}(k) &= 
        \begin{cases}
    		1 & \text{if $A_n(k) = m$ and $F_m(k)=0$}\\
    		0 & \text{otherwise}.
    	\end{cases} 
\end{align}

Since each device can only transmit on one resource at each time slot, the indicator whether the transmission on any resource for the device $n$ has been successful or not, can be written as $G_n(k) = \sum_m G_{n,m}(k) \in \{0,1\}$ and similarly for collision we can write $C_n(k) = \sum_m C_{n,m}(k) \in \{0,1\}$.

Furthermore, we can define matrices with $n$ rows and $m$ columns for success and collision events respectively as,
\begin{align}
    \boldsymbol{G} = \big( G_{n,m} \big) \in \mathcal{B}^{n \times m},
\end{align}
and 
\begin{align}
    \boldsymbol{C} = \big( C_{n,m} \big) \in \mathcal{B}^{n \times m}.
\end{align}

We assume that each user keeps a record of its previous actions, feedback and its current buffer state $B_n(k)$ up to $h$ past instants, where we refer to $h$ as the \emph{history length}. Therefore, the tuple
\begin{align}\label{eq:history_user_n}
    S_n(k)  =  \big( A_n(k-h), \dotso, A_n(k-1), & F(k-h), \dotso, \\ 
    &  F(k-1), B_n(k)\big) \nonumber
\end{align}
is referred to as the \emph{local history} or the \emph{state} of user $n$ at time $k$, and 
$S(k) = \big(S_1(k),\dotsc,S_N(k)\big)$
is the \emph{global history} of the system.

The feedback signal $F(k)$ is only recorded by the devices that are active at time $k-1$. If a new device becomes active at time $k$, its state is initialized with $A_n(k-1)=0$, and $F_m(k-1) = 0, \forall m \in \mathcal{A}$ for its local history $S_n(k)$. The memory is initialized with zero values. Moreover, we set zero values for the time a device has been inactive if the time of inactivity is smaller than the history size. 


\begin{definition}
A policy or access scheme of user $n$ at time slot $k$, is a mapping from $S_n(k)$ to a conditional probability mass function $\pi_n(\cdot|S_n(k))$ over the action space $\{0, 1\}$. We consider a distributed setting in which there is no coordination or message exchange between users for the channel access. 
Each new action $A_n(k) \in \{0, 1\}$ is drawn at random from $\pi_n( \cdot |S_n(k))$ as follows:
\begin{equation}
    \mathrm{Pr}\bigl\{ A_n(k) = a \bigm| S_n(k) = s \bigr\}
    = \pi_n(a|s).
\end{equation}
\end{definition}

We are interested in developing a distributed transmission policy for slotted RA that can effectively adapt to changes in the traffic arrivals and provide better performance in terms of throughput, latency and fairness than the baseline reference schemes. 
We consider \gls{eb} policies as our baseline schemes. More specifically, we use \gls{beb} when the value of backoff factor is $2$, which has been used in IEEE 802.11 and IEEE 802.3 standards. 

\subsection{PERFORMANCE METRICS}

\subsubsection{Throughput}
The channel throughput is defined as the average number of packets that are successfully transmitted from all the devices divided by the total number of \glspl{prb}, over a time window of size $K$. For the finite time horizon $K$ and for $M$ orthogonal resources, the average throughput of the system is defined as
\begin{align}\label{eq:throughput_m}
    T = \frac{1}{M K}\sum_{k=1}^K \sum_{m=1}^M \sum_{n \in \mathcal{N}} G_{n,m}(k).
\end{align}
where $G_{n,m}(k)$ refers to the success event over channel $m$ and $T \in [0, 1]$. 

\subsubsection{Age of Packets}
The \gls{aop}\footnote{This metric has a different connotation to age of information (AoI).} of device $n$, denoted as $w_n(k)$, grows linearly with time if a packet stays in the buffer of the device, and it is reset to $0$ if the packet is transmitted successfully. Specifically, we assume that $w_n(1)=0$, and the \gls{aop} $w_n(k)$ evolves over time as follows:
\begin{align}
    w_n(k) &=
    \begin{cases}
        0  &\text{if } B_n(k) =0 \\
        w_n(k-1) + 1 & \text{otherwise.}
    \end{cases}
\end{align}\label{eq:aop}

The average \gls{aop} for user $n$ after a time span of $K$ time slots is given by
\begin{align}\label{eq:aop_n}
    \Delta_n = \frac{1}{K}\sum_{k=1}^{K} w_n(k)
\end{align}
and the average \gls{aop} of the overall system by $\Delta = \frac{1}{N} \sum_n \Delta_n$. 
 Since techniques such as \gls{eb} incur \emph{capture effect} \cite{beb_analysis} where a transmitting device keeps transmitting on the channel for some time, introducing short-term unfairness. In this work, we use the average \gls{aop} to measure fairness as well as the average delay budget of the packets. A higher \gls{aop} means the scheme is more unfair and has the higher delay and vice versa.
 \begin{figure}
    \centering
        \begin{subfigure}{.24\textwidth}
        \includegraphics[width=110px, height=105px]{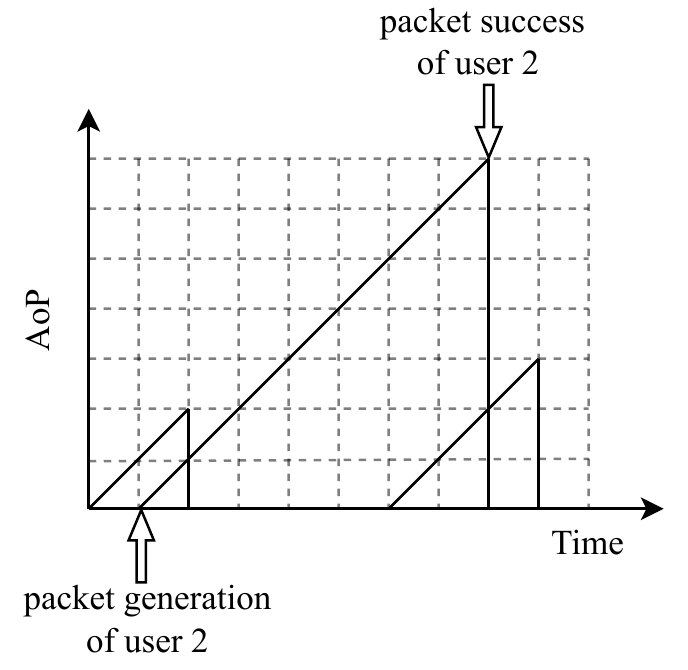}
        \caption{Unfair Scheme}
        \label{fig:aop_unfair}
        \end{subfigure}
        \begin{subfigure}{.24\textwidth}
        \includegraphics[width=125px, height=120px]{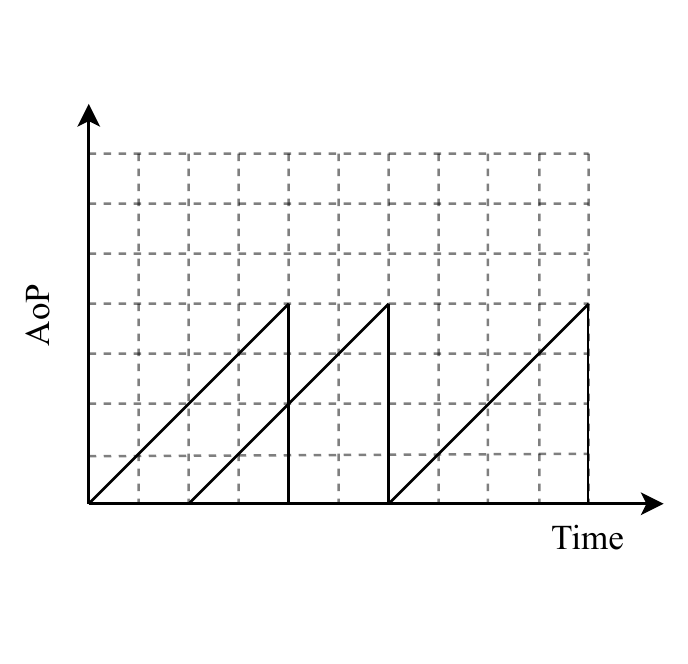}
        \caption{Fair Scheme}
        \label{fig:aop_fair}
        \end{subfigure}

    \caption{An example of using \gls{aop} for fairness.}
    \label{fig:aop_ill}
\end{figure}
To illustrate the concept of fairness with \gls{aop}, let us consider an example of $3$ users where each user generates a single packet that is successfully transmitted within $K=10$ time slots as depicted in Fig. \ref{fig:aop_ill}. The \emph{average} delay (number of time slots taken by each user to transmit their packet is same for both fair and unfair schemes, which is $4$ time slots. However, the scheme shown in Fig. \ref{fig:aop_unfair} is clearly not fair since user $2$ takes much more time slots to send its packet as compared to the other two users. This short-term unfairness can be captured with the average \gls{aop} and we see that the average for the scheme in Fig. \ref{fig:aop_unfair} is higher i.e., $1.23$) than the one shown in Fig. \ref{fig:aop_fair} which is $1.00$.

\section{TRAFFIC MODELS}\label{sec:correlated_model}
\subsection{REGULAR TRAFFIC MODEL}
In traditional \gls{ra} schemes, the activation of \glspl{mtd} and the traffic arrival for each user is usually modeled by an independent process. We call such traffic arrival as \emph{regular} traffic arrival. Each device follows independent Bernoulli process with average arrival rate $\lambda_n$ to generate packets in regular traffic model. The average arrival rate for the system can then be written as 
\begin{align}
    \lambda = \sum_n \lambda_n
\end{align}

However, for \gls{mtc}, it is highly likely that some devices are correlated in terms of activation, i.e., some devices observe the same physical phenomenon and activate together. For instance, in industrial fault detection or fleet management, some \glspl{mtd} are highly likely to transmit at the same time due to the activation of certain event. For intance, in flood or quake detection or land sliding, there is a high probability that devices closer to the event will start transmitting at once. Several recent works have used a correlated activity model to design access schemes for \gls{mtc} \cite{alberto_21, Anders_18, zheng_22, federico_21}.  Therefore, the assumption of independent traffic arrival is not valid in this case. Moreover, apart from correlated \gls{ed} device activity, each \gls{mtd} also follows regular traffic model \cite{Thomsen17}.

\subsection{CORRELATED TRAFFIC MODEL}

\begin{figure}
    \centering
    \includegraphics[width=220px, height=175px]{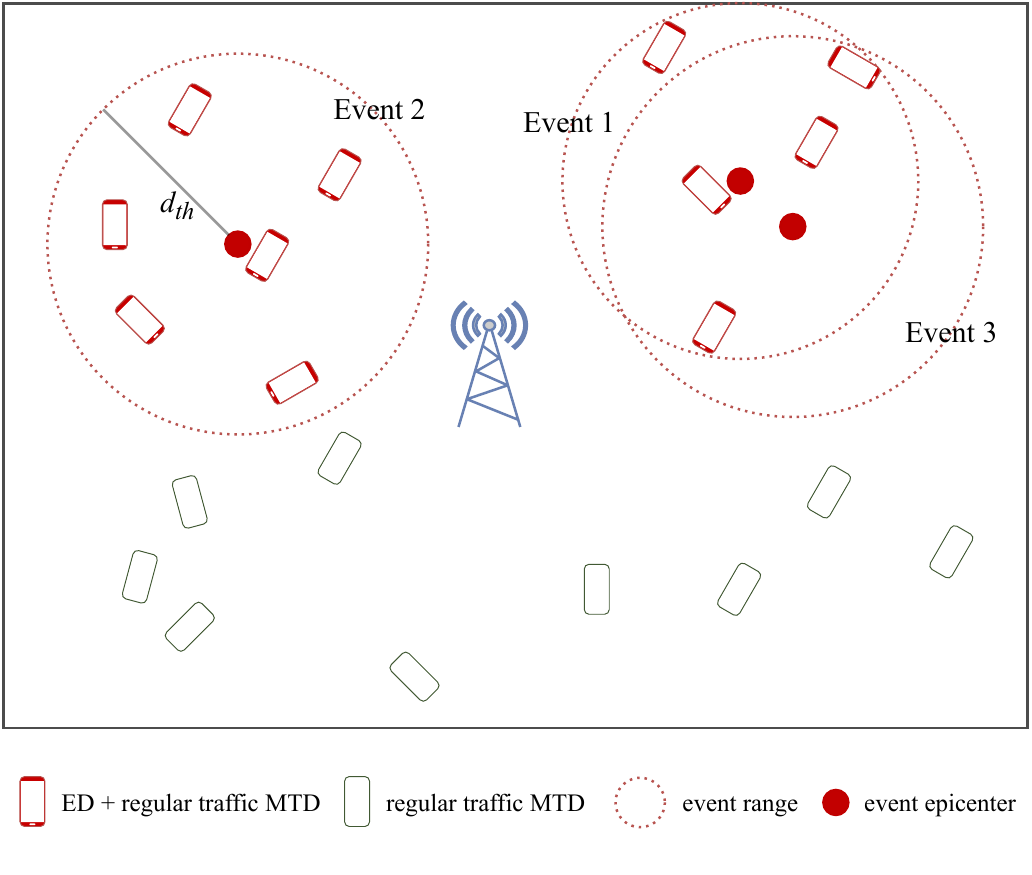}
    \caption{MTC Network depicting $N=20$ \glspl{mtd} uniformly distributed in a rectangular area with $L=3$ event epicenters. \Glspl{mtd} follow regular traffic and those within the range of the event epicenter follow both regular and \gls{ed} traffic.}
    \label{fig:ed_reg_traffic}
\end{figure}

The correlated traffic model is a mix of the  \emph{regular} traffic and \emph{\gls{ed}} traffic or \emph{alarm} traffic as depicted in Fig. \ref{fig:ed_reg_traffic}. 
We assume that the regular traffic generation for each \gls{mtd} follows an independent random process such as Bernoulli process. Similarly, the \gls{ed} traffic also follows a random process on top of the regular traffic arrival process. For \gls{ed} traffic, certain devices are strongly correlated in space and time and the \gls{ed} traffic generation for such devices is dependent on the occurrence of an event in their vicinity. Regular traffic arrival and \gls{ed} traffic arrival processes are independent of each other. 

To formulate this behaviour, we assume that $N$ \glspl{mtd} are uniformly distributed in a given area. Each \gls{mtd} can either be in a regular state or alarm state, when active. We consider $L$ event epicentres that are scattered randomly and independently across the given area. The location of the devices is represented by $\mathbf{x} \in \mathbb{R}^2$ and the location of the epicenter of the events is denoted by $\mathbf{y} \in \mathbb{R}^2$. We assume that all \glspl{mtd} are stationary and fixed to their locations or exhibit very low mobility. 

Let $E_{\mathbf{x}\mathbf{y}}$ denote the event when a device $n \in \mathcal{N}$ at location $\mathbf{x}$ is triggered into alarm or \gls{ed} mode by the activation of an event with its epicenter at location $\mathbf{y}$. Let $\bar{E}_{\mathbf{x}\mathbf{y}}$ be the complement of $E_{\mathbf{x}\mathbf{y}}$ and   $p_{\mathbf{x}\mathbf{y}}$ denotes the probability of a device $n$ at location $\mathbf{x}$ being triggered into alarm mode by the activation of event at location $\mathbf{y}$.
Moreover, we define the probability of device at location $\mathbf{x}$ being in \gls{ed} mode is $p_{\mathbf{x}}$. We write,
\begin{align}
    p_{\mathbf{x}} & = \mathrm{Pr} \big[\text{At least one event triggers \gls{mtd} at } \mathbf{x} \big] \nonumber \\ 
    & = 1 - \mathrm{Pr} \big[\text{No event triggers \gls{mtd} at } \mathbf{x} \big] \nonumber \\ 
    & = 1 - \prod_{\mathbf{y} \in \mathcal{L}} \mathrm{Pr}\big[\bar{E}_{\mathbf{x}\mathbf{y}}\big] \\
    & = 1 - \prod_{\mathbf{y} \in \mathcal{L}} (1 - p_{\mathbf{x}\mathbf{y}}),
\end{align}
where we have assumed that events are triggered independently of each other.

Therefore, for the correlated traffic arrival model, each \gls{mtd} can either be at \emph{alarm} (\gls{ed}) state or \emph{regular} state for a given time slot $k$. We denote by $V_{\mathbf{x}}$ the state of device at location $\mathbf{x}$ and we model the states at each time slot $k$ by i.i.d Bernoulli random variable as,
\begin{align}
    V_{\mathbf{x}}(k) = 
    \begin{cases}
    \text{Regular with prob. } 1 - p_{\mathbf{x}} \\ 
    \text{Alarm with prob. } p_{\mathbf{x}}
    \end{cases}
\end{align}
Moreover, we define with $p$ the probability of an event being active at location $\mathbf{y}$. We assume that each event at epicenter $\mathbf{y}$ triggers a subset $ \mathcal{N}_\mathbf{y} \subseteq \{1,\dotsc, N\}$ of devices. The probability of a device $n$ going into alarm mode depends on the distance of the device from the epicenter $\mathbf{y}$ of the event. Furthermore, each \gls{mtd} $n$ can sense and report multiple events but at any given time, we assume that it can report about only one event. The \glspl{mtd} are unaware of the actions, and events sensed by other devices.

\renewcommand{\baselinestretch}{0.8}

\begin{table*}
	\caption{Suitability of some popular MARL for designing \gls{ra} schemes for \gls{mtc} systems}
	\centering
	\begin{tabular}{|p{2.5cm}|p{7.5cm}|p{6cm}|} 
	\hline
	\textbf{MARL Algorithm} & \textbf{Features} & \textbf{Limitations}  \\ \hline \hline
	\makecell[l]{ \\
		\gls{dqn} \cite{mnih2015humanlevel} \\
		and variants
	} &
	Can be used as independent Q learning (IQL) or centralized learning using \gls{ps} by extending single agent network to multiple agent. & Has convergence issues and it is nearly impossible to learn optimal policy for a large number of agents with \gls{ps}. \vspace{0.1cm} \\
	
        \hline 
	\makecell[l]{
		\acrshort{maddpg} \cite{maddpg_LoweWTHAM17} \\
		\acrshort{mappo} \cite{mappo_chao2021}
	} & 
	\begin{minipage}{\linewidth}
        \vspace{\baselineskip}
	\begin{itemize}[leftmargin=4mm]
		\item They have a shared critic and local actors.
		\item \acrshort{mappo} is on-policy and \acrshort{maddpg} is off-policy for \gls{ctde} method. 
		\item Both circumvent the challenge of non-Markovian and non-stationary environments during learning. 
		\item Stabilize learning, due to reduced variance in the value function estimates.
	\end{itemize} 
	\end{minipage}
	\vspace{1mm}
	&
	\multirow{2}{*}[5mm]{
	\begin{minipage}{\linewidth}
	\begin{itemize}[leftmargin=4mm]
		\item Need centralized critic which is not scalable to a large number of
		agents. 
		\item The state-space dimensionality grows exponentially for the critic as the number of agents increases.
		\item Most critically, the accumulated noises by the exploratory actions of other agents make the Q-function learning no longer feasible \cite{qmix}. 
	\end{itemize}
	\end{minipage}
	}
	\\
	\cline{1-2}
	\makecell[l]{
		\acrshort{coma} \cite{coma} 
	} &
	\begin{minipage}{\linewidth}
        \vspace{\baselineskip}
        
	\begin{itemize}[leftmargin=4mm]
		\item Tackles the credit assignment problem.
		\item Calculates the advantage function which is able to marginalize a single agent’s actions while keeping others fixed. 
	\end{itemize}
        \vspace{0.1cm}
	\end{minipage}
	& \\
	\hline
	\makecell[l]{
		QMIX \cite{qmix} \\
		 \gls{vdn} \cite{vdn}
	} &
	\begin{minipage}{\linewidth}
	\begin{itemize}[leftmargin=4mm]
		\item Lie between \acrshort{coma} and IQL. 
		\item Better scalability than centralized critic methods.
		\item Each agent has its own network and then mixing network takes the Q-values of agents to measure $Q_{tot}$.
		\item  QMIX uses NNs to make the function monotonic whilst \gls{vdn} is the linear combination of the Q values of each agent.
		\item Both of them make use of \glspl{rnn}.
	\end{itemize}
	\end{minipage}
	& 
	\begin{minipage}{\linewidth}
        \vspace{\baselineskip}
	\begin{itemize}[leftmargin=4mm]
		\item Scalability to a massive number of agents.
		\item Different policy for each agent and it can also be applied to homogeneous agents with \gls{ps}.
		\item The information might be lost in this way of centralized training.
		\item It is not sure whether the training with \gls{ps} without using agent IDs will result in better policy than \gls{dqn} or not.
	\end{itemize}
        \vspace{0.1cm}
	\end{minipage}
	\\ \hline 
	\makecell[l]{\\
		Mean Field MARL \\
		(MF-MARL) \cite{mf_marl}
	}
	& Good scalability results and and it considers the effects of neighbouring agents for each agent to estimate its value function & Requires communication between agents; each agent has its own
	policy to learn. \vspace{0.1cm} \\
	\hline 
	\vspace{\baselineskip} Multi-Actor-Attention-Critic (MAAC) \cite{maac_iqbal} 
        
	& \vspace{\baselineskip} Uses attention mechanism to incorporate the effects of important agents whom actions are given more consideration than others 
	& \vspace{\baselineskip} Requires communication among agents, scales better than \acrshort{maddpg} but their results are only for a small number agents. \\
	\hline
	Soft Actor Critic \cite{soft_ac} & 
	\begin{minipage}{\linewidth}
        \vspace{\baselineskip}
	\begin{itemize}[leftmargin=4mm]
		\item Extension of AC methods. Uses the notion of entropy to encourage exploration and to avoid converging to non-optimal policies. 
		\item Can be used for multiagent system with \gls{ps}, local actors and local critics.
	\end{itemize}
	\end{minipage}
	& 
	\begin{minipage}{\linewidth}
        \vspace{\baselineskip}
	\begin{itemize}[leftmargin=4mm]
		\item Has convergence issues just like \gls{dqn}.
		\item The performance is not known with centralized training, i.e., centralized critic and centralized actor, an extension of single agent system just like \gls{dqn}. \vspace{0.1cm}
	\end{itemize} 
	\end{minipage} \newline
	\\ 
	\hline
	\end{tabular}
	\label{tab:marl_suitability_mtc}
\end{table*}
\renewcommand{\baselinestretch}{1.0}

\section{SUITABILITY OF MARL FOR RA SCHEME DESIGN}\label{sec:marl_suitability}
To design a \gls{ra} policy for the multiuser \gls{mtc} environment, it is important to consider the suitability of \gls{marl} algorithms in general for scalability and in particular for the specific characteristics of \gls{mtc} system. There exists a large body of the literature for medium access using \gls{rl} in wireless networks but only a few address the issues of scalability (e.g., \cite{mateus_vtc22} and our recent work \cite{jadoon_vtc}). The distributed multiple schemes designed with \gls{marl} do have scalability challenges and this is even further exacerbated by the limitations of the \gls{mmtc}. The \gls{mtc} system presents the following challenges for \gls{marl} algorithms in designing a distributed \gls{ra} policy:
\begin{enumerate}
    \item Since the \glspl{mtd} are low-powered and low-complexity devices that are battery-operated, it is not feasible to perform learning on the devices and therefore, \acrfull{ctde} method is required for learning.
    \item Usually, \gls{ps} method is used for homogeneous agents, and each agent's ID is used in the observation (state) vector to distinguish between agents. The homogeneous agents are those that have the same state-space and action-space. In our system model, since the devices\footnote{the terms agent, device and \gls{mtd} are used interchangeably throughout the paper.} have variable sleep cycles and they should be able to join/leave the network, it is not feasible to use agent identification. Therefore, we require each agent to have a single policy using \gls{ps} but at the same time, without using agent IDs. 
    \item Any communication to exchange channel or device information between \glspl{mtd} is not energy efficient as it will drain the battery of the devices. Therefore, the devices do not communicate with each other and they do not know the actions taken by the other devices.
\end{enumerate}

In Table \ref{tab:marl_suitability_mtc}, we provide the comparison of some popular \gls{marl} algorithms and their suitability to the proposed system model. Even though there are several \gls{marl} algorithms found in the literature, we have given the comparison of some well-known algorithms that employ the \gls{ctde} method to learn policies. The aim of this comparison is not to provide an exhaustive survey of \gls{marl} algorithms but to make a case of using a specific algorithm over the others for the proposed system model. Interested readers are referred to \cite{Vouros22}, a recent review paper of different \gls{marl} algorithms addressing scalability challenges. Moreover, we are focusing on \gls{drl} algorithms only.

Standard \gls{dqn} \cite{mnih2015humanlevel} and actor-critic algorithms are extended to multiple agents using \gls{ps} for homogeneous agents in \cite{gupta_2017}. This method scales well to a large number of agents but it does not exploit any advantages of centralisation. Moreover, without using IDs of agents and without any cooperation between agents, we have observed in our simulations that these algorithms are not able provide better policies and they have convergence issues. However, this might be the issue for most of the centralized approaches. Our recent works \cite{jadoon_wcnc22} and \cite{jadoon_vtc} use the \gls{dqn} with \gls{ps} and in \cite{jadoon_vtc}, we provided results for up to 500 devices for the bursty traffic. Similarly, just like the \gls{dqn} can be extended to multiple agents using \gls{ps}, one can also use Soft Actor Critic method \cite{soft_ac} with a local actor and local critic that is shared among all users, where users' individual observation is used to update the actor and critic at each time step for all the users.

Other popular choices for \gls{marl} are \gls{maddpg} \cite{maddpg_LoweWTHAM17} and the recently proposed \gls{mappo} \cite{mappo_chao2021}. In both \gls{maddpg} and \gls{mappo}, the critic network has a global view of the system, which is only applied during the training phase and actor networks are employed for each agent. A major bottleneck of these algorithms is the scalability due to the shared critic network, even if the \gls{ps} is considered. Since the shared critic network has the observation space of all the agents, the size of the observation space will grow exponentially with the number of agents. Moreover, in a network where the number of agents changes with time,  it is inefficient to use the state-space of all agents at the centralized critic. \gls{coma} \cite{coma} has a similar issue for scaling to a higher number of agents because a shared critic is used in it as well, just like \gls{maddpg} and \gls{mappo}. 

Some algorithms such as \gls{mfmarl} \cite{mf_marl} and \gls{maac} \cite{maac_iqbal} show good scalability results but the major issue in these algorithms is that they require communication between the agents, which is not practical for our system model. Furthermore, the results for these algorithms are shown for a moderate number agents. 

\Gls{vdn} \cite{vdn} and QMIX \cite{qmix} are both  for cooperative multi-agent learning in which joint action-values $Q_{tot}$ are estimated from Q-values of individual agents that condition only on local observations. One of the main differences between these algorithms is that in \gls{vdn}, the $Q_{tot}$ is calculated as a linear combination of the Q-values of each agent, while QMIX employs a network that can compute $Q_{tot}$ as complex non-linear combination of individual Q-values. This way of learning also provides better scalability as compared to \gls{maddpg} and \gls{mappo}. QTRAN \cite{qtran_19} improves upon both \gls{vdn} and QMIX and provides more general form of factorization but it falls under the same category as \gls{vdn} and QMIX. For these reasons, we will focus on the \gls{vdn} and QMIX algorithms in our simulations.

\section{RL ENVIRONMENT AND MARL ALGORITHMS}

\subsection{THE ENVIRONMENT}
We consider shared \glspl{prb} where each agent interacts with the resources by taking an action and receiving a common feedback $F(k)$ as observation. The reward $R_n(k)$ is then calculated by each agent. The action space of each agent is $\mathcal{A}$ and each device can either transmit on channel $m$, i.e., $A_n(k) = m$ or it can stay silent, i.e., $A_n(k)=0$. The state of each device is the history tuple $S_n(k)$ defined in \eqref{eq:history_user_n}. Let $R_n(k) \in \mathbb{R}$ be the \emph{immediate} reward that user $n$ obtains at the end of time slot $k$. The reward depends on the agent $n$ action $A_n(k)$ and other agents' actions $A_{n'}(k)$, $n' \neq n$. The accumulated discounted reward for user $n$ is defined as
\begin{align}
    \mathcal{R}_n(k) = \sum\limits_{k'=0}^{\infty}\gamma^{k'} R_n(k+k'+1),
\end{align}
where $\gamma \in [0,1)$ is a discount factor.
The reward function to maximize the packet success rate is defined as, 

\begin{align}\label{eq:rwd_thput}
    R_n(k) = \sum_{i=1}^N \sum_{m=1}^M & G_{i,m}(k), \quad \forall i \in \mathcal{N}, \forall m \in \mathcal{M}
\end{align}
The summation sign over agents shows that the reward  is global, i.e., all agents receive the same reward, which indicates that the agents are fully cooperative. The environment is partially observable as each agent is unaware of the actions taken by the other user,

At each time slot $k$, each agent $n$ obtains the feedback $F(k)$ from the receiver, updates its history and then feeds $S_n(k)$ to the proposed algorithm, whose output are the Q-values for all the available actions. Each agent $n$ follows the policy $\pi$ by drawing an action $A_n(k)$ from the following Boltzmann distribution,
\begin{align}
    \pi(a|s)
    &= \frac{e^{ Q(a,s)/\tau}} {\sum\nolimits_{\tilde{a} \in \{0,\dots,M\}} e^{Q(\tilde{a},s)/\tau}},
    \qquad \forall a \in \{0, \dots, M\},
    \label{eq:strategy}
\end{align}
where $0 < \tau < \infty $ is the temperature parameter which is used for \emph{exploration}. We decrease the value of $\tau$ to $0$ gradually to make the agent more greedy.s

\subsection{DEEP Q-NETWORK (DQN)}
\Gls{dqn} represents the action-value function using a neural network that are characterized by parameter $\theta$. In double \gls{dqn}, a target network is also used which is parameterized by $\theta^-$ that are periodically copied from $\theta$ during training. The \gls{dqn} and its variants use a \textit{replay buffer} to store the transitions $(s,a,r,s')$, where $s$ is the actual state, $s'$ is the next state that is observed after taking action $a$ and receiving reward $r$. The learning updates are applied on the experience samples $(s,a,r,s') \sim U(\mathcal{D})$, that are drawn at random with uniform distribution as mini-batches of size $z$ from $\mathcal{D}$ and by minimizing the following loss function,
\begin{align}\label{eq:dqn_loss}
    L(\theta) = \sum_{i=1}^z \Big[ \big(y_i^{DQN}
    & - Q(a_i, s_i; \theta_i)\big)^2 \Big], 
\end{align}
where $y_i^{DQN} =r + \gamma \max_{a'} Q(a', s'; \theta_i^-)$ is the target value for the $i^\text{th}$ iteration. 

Since the environment is partially observable, the agents can benefit from using \gls{rnn} such as \gls{gru} that can facilitate learning from previous history. A \gls{dqn} making use of \gls{rnn} is referred to as \gls{drqn}.

\subsection{VALUE DECOMPOSITION NETWORKS (VDN)}
\Gls{vdn} \cite{vdn} take advantage of centralization and aim to learn the joint action value function $Q_{tot}(\boldsymbol{s}, \boldsymbol{a})$, a linear value decomposition from the team reward signal, where $\boldsymbol{s}$ is the joint observation of the agents and $\boldsymbol{a}$ is the joint action of agents. The \gls{vdn} algorithm decomposes the $Q_{tot}$ as the linear combination of the individual Q-values of each agent, i.e.,
\begin{align}\label{eq:vdn_mixing}
    Q_{tot}(\boldsymbol{a}, \boldsymbol{s}) \approx 
        \sum_{n=1}^N Q(a_n, s_n; \theta_n),
\end{align}
The loss function of the \gls{vdn} algorithm can be calculated in the same way as \gls{dqn}, i.e.,
\begin{align}\label{eq:vdn_loss}
    L(\theta) = \sum_{i=1}^z \Big[ \big(y_i^{tot}
    & - Q_{tot}(\boldsymbol{a}, \boldsymbol{s}; \theta_i)\big)^2 \Big], 
\end{align}
where 
\begin{align}\label{eq:y_total}
    y_i^{tot} =r + \gamma \max_{\boldsymbol{a}'} Q(\boldsymbol{a}', \boldsymbol{s}'; \theta_i^-)    
\end{align}
is the target value for the iteration $i$. 

In this way, each agent performs an action selection locally based on its own learned Q-value in a decentralized manner. Moreover, the \gls{vdn} method employs \gls{rnn} or \gls{drqn} to calculate Q-values for each agent. 

\subsection{QMIX}
The QMIX \cite{qmix} algorithm improves the \gls{vdn} and it can represents much richer class of action-value functions. QMIX applies the following constraint on the relationship of $Q_{tot}$ and each individual action-value $Q_a$,
\begin{align}\label{eq:qmix_constraint}
    \frac{\partial Q_{tot}}{\partial Q_a} \geq 0 \quad \forall a,
\end{align}
to ensure that mixing network has positive weights. Intuitively, it shows that if the weights of individual value function $Q_a$ are negative, less weightage is given to that agent for cooperation. Moreover, as opposed to \gls{vdn}, $Q_{tot}$ is calculated in a complex non-linear way. QMIX uses a separate feed-forward neural network as a mixing network that takes individual agents' outputs and mixes them monotonically to produce $Q_{tot}$ to enforce the constraint in \eqref{eq:qmix_constraint} \cite{qmix}. The weights of the mixing network are produced by a separate \emph{hypernetwork} to ensure that they are non-negative. The loss function is calculated in the same way as given in \eqref{eq:vdn_loss}.

\begin{algorithm}[t]
\caption{Training Phase of the Proposed Algorithm}\label{alg:alg1}
\begin{algorithmic}
\STATE 
\STATE \textbf{Define} $N, \tau, \gamma \in [0, 1], \lambda_n, \forall 
n \in \mathcal{N}, h$ and $K$
\STATE \textbf{Initialize} $S_n = 0, B_n = 0,  \forall 
n \in \mathcal{N}$
\STATE \hspace{0.5cm}$ \textbf{for} \text{ each episode} \textbf{ do} $
\STATE \hspace{1.0cm}$ \textbf{for} \text{ each time slot } k=1,\dots,K \textbf{ do} $
\STATE \hspace{1.5cm} \textsc{Generate} traffic for all \glspl{mtd}, i.e., $\tilde{B}_n\sim$ \STATE \hspace{1.5cm} $Bernoulli(\lambda_n)$ for regular traffic, and $\tilde{B}_n\sim$ \STATE \hspace{1.5cm} $ Bernoulli(p)$ for \gls{ed} traffic, $\forall n \in \mathcal{N}$ 
\STATE \hspace{1.5cm} \textsc{Update} buffer $B_n = \min(1, \tilde{B}_n)$
\STATE \hspace{1.5cm}$ \textbf{for} \text{ each agent/\gls{mtd} }  n=1,\dots,N \textbf{ do}$
\STATE \hspace{2.0cm} Observe input $S_n$ and feed it to \gls{drqn} 
\STATE \hspace{2.0cm} Generate the estimate of $Q_a, \forall a \in \mathcal{A} $
\STATE \hspace{2.0cm} Choose action according to \eqref{eq:strategy}
\STATE \hspace{2.0cm} Receive feedback $F(k)$ from the receiver
\STATE \hspace{2.0cm} Obtain reward $R_n$ according to \eqref{eq:rwd_thput}
\STATE \hspace{2.0cm} Update buffer $B_n$
\STATE \hspace{2.0cm} Observe the next state $S_n$
\STATE \hspace{1.5cm} $\mathbf{end}$
\STATE \hspace{1.5cm} \textsc{Store} $(S, \mathbf{A}, \mathbf{R}, S' )$ in the replay buffer $\mathcal{D}$, \STATE \hspace{1.5cm} where $\mathbf{A} = (a_1, \dots, a_N)$, $\mathbf{R} = (R_1, \dots, R_N)$
\STATE \hspace{1.5cm} \textsc{Set} $S \leftarrow S'$
\STATE \hspace{1.5cm} \textsc{Sample} a minibatch $(S^i, \mathbf{A}^i,\mathbf{R}^i, S'^i)$ from $\mathcal{D}$
\STATE \hspace{1.5cm} \textsc{Calculate} $y^{tot}$ using \eqref{eq:y_total}
\STATE \hspace{1.5cm} $\mathbf{if}$ mixer $=$ \gls{vdn}:
\STATE \hspace{2.0cm} \textsc{Calculate} $Q_{tot}$ using \eqref{eq:vdn_mixing}
\STATE \hspace{1.5cm} $\mathbf{else if}$ mixer $=$ QMIX:
\STATE \hspace{2.0cm} \textsc{Calculate} $Q_{tot}$ using QMIXer \cite{qmix}
 
\STATE \hspace{1.5cm} $\mathbf{for}$ every $K^{\theta}$ time slots:
\STATE \hspace{2.0cm} {\textsc{Update}} $\theta^- \leftarrow \theta$
\STATE \hspace{1.5cm} $\mathbf{for}$ every $K^{\beta}$ time slots:
\STATE \hspace{2.0cm} {\textsc{Update}} $\beta$
\STATE \hspace{1.0cm} $\mathbf{end}$
\STATE \hspace{0.5cm} $\mathbf{end}$
\end{algorithmic}
\label{alg1}
\end{algorithm}
\begin{table}
\centering
\caption{Simulation Parameters}
\renewcommand{\arraystretch}{1.0}
    \begin{tabular}{|l|l|}
    \hline
    \textbf{Parameter}                             & \textbf{Value}    
     \\ \hline \hline
    $\tau$       & 200 -- 0.1      \\ \hline
    $\lambda_n$       & 0.3 (regular) and 0.015 (correlated)                                                 \\ \hline
    Total Episodes         & $60$                                            \\ \hline
    History size $h$                      & 5 if RNN else 1                                        \\ \hline
    Learning rate  & $10^{-4}$           \\ \hline
    Batch size         & 32 \\ \hline
    $d_{th}$         & 0.3 \\ \hline
    \end{tabular}
\label{tab:simulation_params}
\end{table}
\section{SIMULATION RESULTS AND DISCUSSION}
In the following experiments, we use the \gls{vdn} \cite{vdn} method to learn \gls{ra} policies for different values of $N$ and $M$. We use the neural network with two layers of size $256$ and $64$ units before the final layer of size $M$, and when \gls{rnn} is used, a \gls{gru} layer of $64$ neurons is added after the first layer. The parameters used during the training of the network are presented in Table \ref{tab:simulation_params}. In all the experiments, we use experience replay to accumulate each agent's experience and the learning is performed with \gls{ctde} method. An agent's ID is one-hot encoded vector that is appended with the observation $S_n(k)$ of each agent. We will first provide results for regular traffic in which we also compare the results for different \gls{marl} algorithms such as QMIX and \gls{drqn} with \gls{vdn} and then we present our results for the correlated traffic arrival model. 
\subsection{RESULTS FOR REGULAR TRAFFIC}
For regular traffic, we employ two ways of training all algorithms: \begin{enumerate*}[label=(\roman*)] \item using agents' IDs in the observation vector, which is a common way of training for \gls{ctde}, and \item without using them. \end{enumerate*} We denote IDs $ =0$ as the case when agent IDs are not used and IDs $=1$ as the case when we incorporate agent IDs in the observation space.
\begin{figure}[h]
\centering
    \includegraphics[width=240px, height=170px]{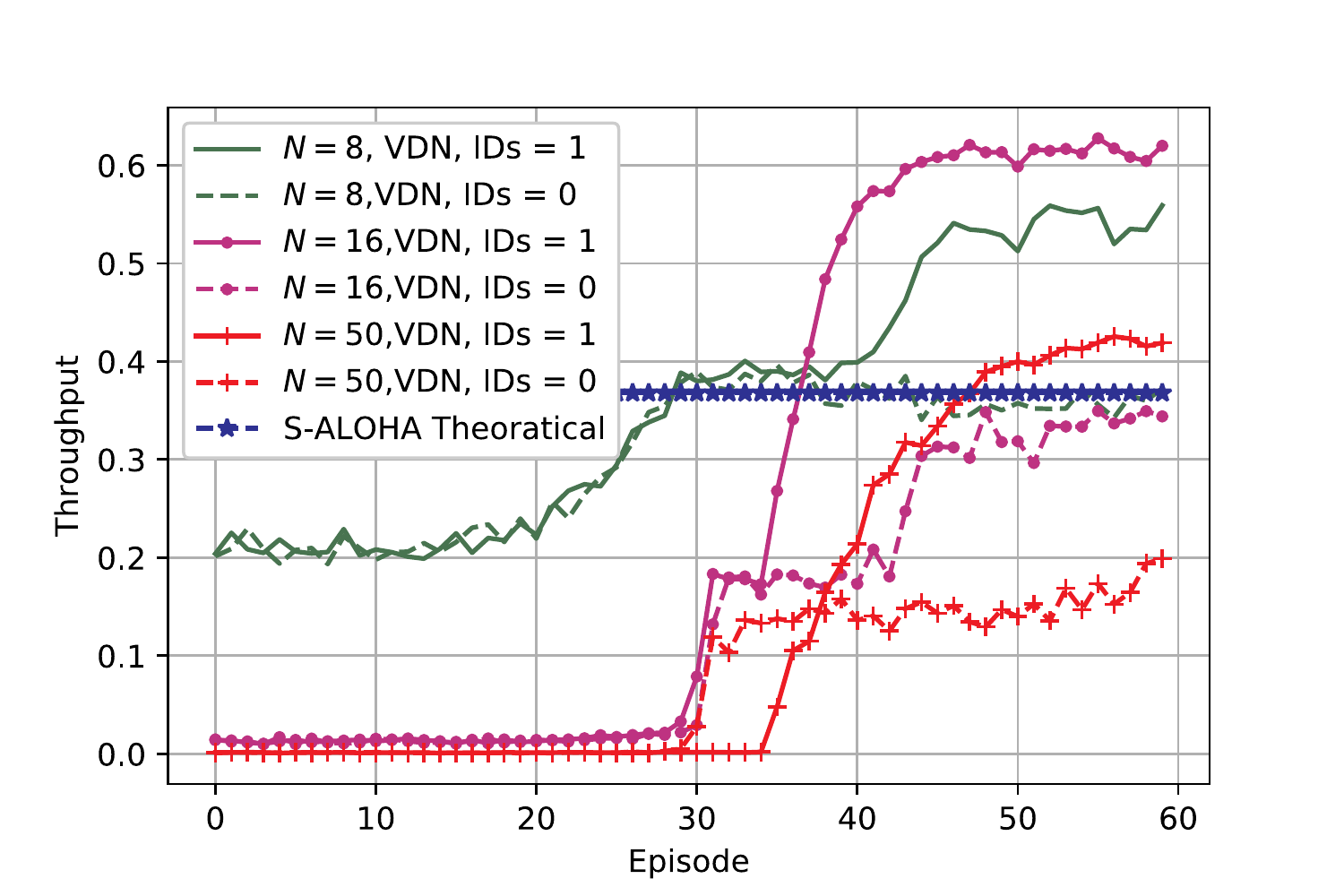}
    \caption{Average throughput during training with \gls{vdn} algorithm for different values of $N$ and to compare the cases when using IDs and not using IDs. The results are for $\lambda_n=0.3$ and $(K, N) = (2000, 8)$, $(K, N) = (3000, 16)$ and $(K, N) = (5000, 50)$.}
    \label{fig:thput_train_allN}
\end{figure}

\begin{table}[t]
\centering
\caption{Average throughput and average \gls{aop} values for the proposed algorithm compared to the \gls{beb} for the learned policies shown in Fig. \ref{fig:thput_train_allN}.}
\renewcommand{\arraystretch}{1.3}
    \begin{tabular}{ |p{0.25cm}|p{0.25cm}|p{0.8cm}|p{0.75cm}|p{0.7cm}||p{0.75cm}|p{0.75cm}|p{0.7cm}| }
     \hline 
     \multicolumn{2}{|c|} {}&
      \multicolumn{3}{c||}{\textbf{Av. Throughput}} &
        \multicolumn{3}{c|}{\textbf{Av. AoP}} \\
     \hline
     $N$ & $M$ & VDN IDs=1 & VDN IDs=0 & BEB & VDN IDs=1 & VDN IDs=0 & BEB \\
    \hline
    $8$ & 2 & $0.56$ & $0.40$ & $0.37$ & $625.7$ & $5.5$ & $162.2$ \\

    $16$ & 2& $0.54$ & $0.386$ & $0.372$ & $1480.4$ & $29.8$ & $519.3$ \\

    $50$ & 5 & $0.44$ & $0.25$ & $0.36$ & $1561.3$ & $44.1$ & $1095.1$ \\ \hline 
    
    \end{tabular}

\label{tab:th_aop}
\end{table}

\begin{figure*}[ht]
    \centering
        \begin{subfigure}{.45\textwidth}
        \includegraphics[width=240px, height=165px]{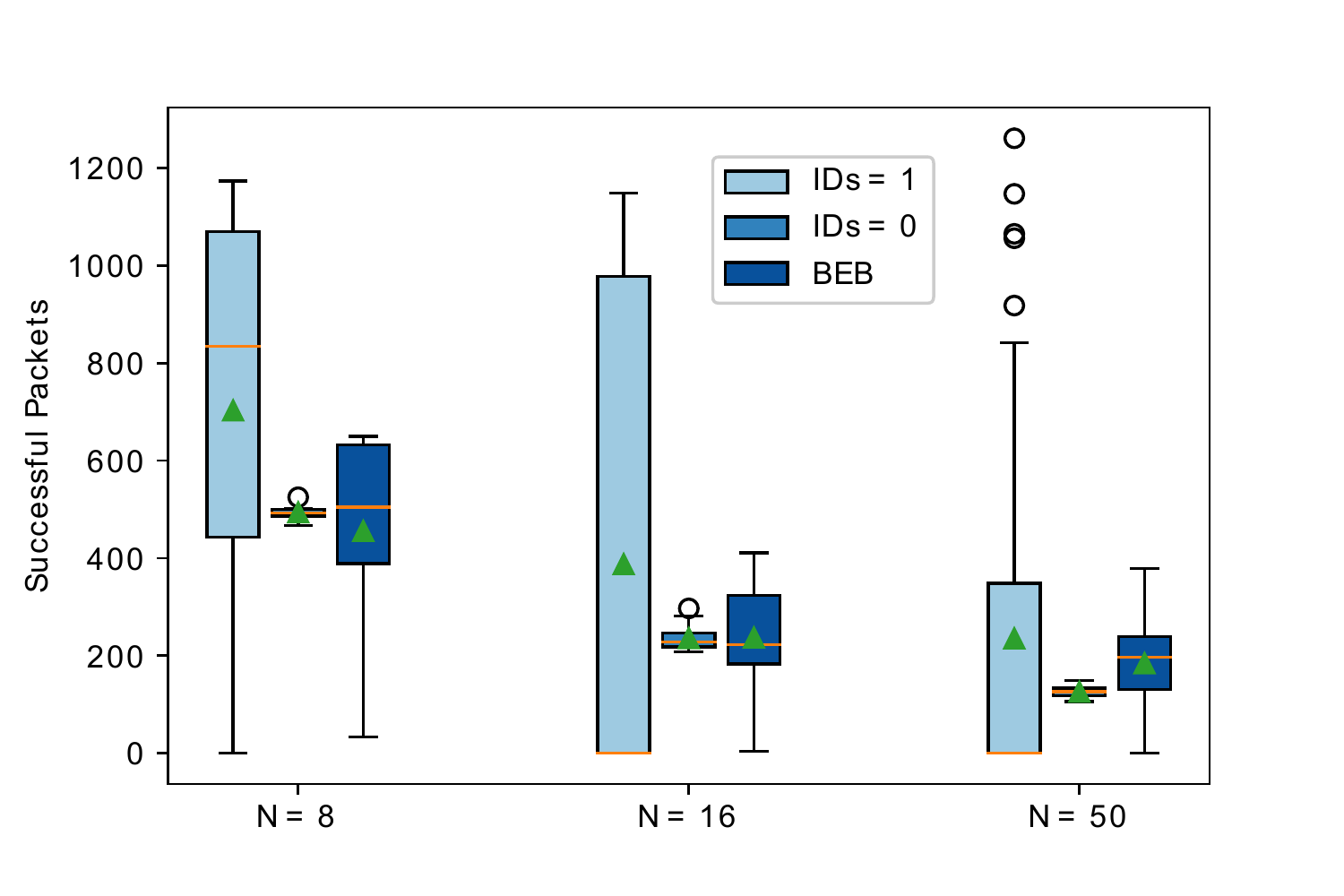}
        \caption{Packets success distribution}
        \label{fig:pkt_dist_all}
        \end{subfigure}
        \begin{subfigure}{.45\textwidth}
        \includegraphics[width=240px, height=165px]{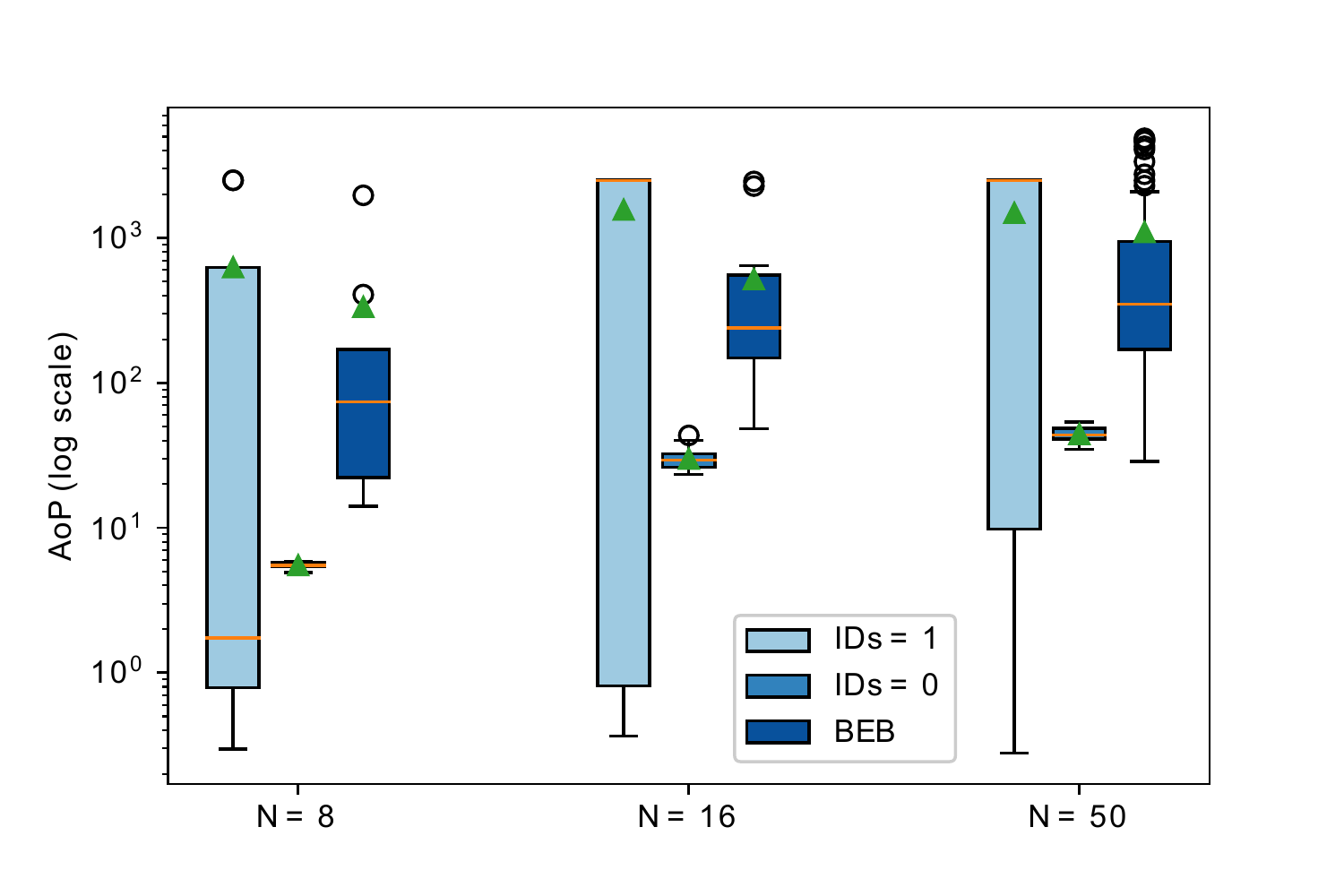}
        \caption{AoP distribution}
        \label{fig:aop_dist_all}
        \end{subfigure}

    \caption{Distribution of successful packets and AoP among \glspl{mtd} for $N = \{8, 16, 50\}$ and $\lambda_n = 0.3$. We used VDN for both IDs $= 1$ and IDs $= 0$ cases. All cases tested for $K = 500$ time slots.}
    \label{fig:dist_all}
\end{figure*}

We show the results in terms of average throughput (normalized reward) and \gls{aop}. The learning process and how average throughput of the system increases for different values of $N$ is shown in Fig. \ref{fig:thput_train_allN}. We use $K= 2000, 3000$, and $5000$ per episode during training for $N=8,16$ and $50$, respectively. We increase $K$ per episode as $N$ grows to allow better learning for each value of $N$. Moreover, $M=2$ is used to $N=8$ and $16$ and $M=5$ is used when $N=50$. The average throughput and average \gls{aop} after testing is shown in Table \ref{tab:th_aop}. Clearly, the case IDs $=1$ outperforms \gls{beb}, slotted ALOHA theoretical throughput (i.e., $1/e$), and the case when IDs$ =0$. The case IDs $=0$ provides much lower average throughput and as the number of devices $N$ grow, the average throughput also decreases which is not surprising because agents decrease the transmit probability as $N$ grows. 

Moreover, we calculate whether the learned policy is fair or not in two different ways as depicted in Fig. \ref{fig:dist_all}. First, we show how many packets per user have been successful as in Fig. \ref{fig:pkt_dist_all} and the second, the \gls{aop} of individual users (which shows both packet delay and fairness) as shown in Fig. \ref{fig:aop_dist_all}. We observe that using IDs incurs a significant unfairness among devices as a subset of \glspl{mtd} are starved out and they never get a chance to send a packet. Surprisingly, no IDs case provides much better fairness among devices. 
It is due to the inherent way the MARL algorithms with \gls{ctde} behave, we see that the case where IDs are omitted, provides us better fairness as compared to the other case. When we use IDs in the state space, devices are only concerned about achieving a better throughput. But when we do not use IDs, there's no such coordination that exists during the centralized training, which could allow the \glspl{mtd} to come up with an unfair consensus. 
 
To understand this, let us take the case of $N= 8$. For \gls{vdn}, IDs $=1$ case, it is clear that there are some devices that have sent $0$ packets and there are a few devices that have sent most of the packets ($95$\textsuperscript{th} percentile is $1070$ and $25$\textsuperscript{th} percentile is around 443). Similarly, in Fig. \ref{fig:aop_dist_all} for $N=8$, the average \gls{aop} value, which is the mean point and the distribution of \gls{aop} among each device shows a significant difference between $75$\textsuperscript{th} percentile ($627$) and $25$\textsuperscript{th} percentile ($0.8$). On the other hand, for \gls{vdn}, IDs $=0$ case, the number of successful packets sent by each device are around the mean value ($493$) for all percentiles, i.e., $75$\textsuperscript{th} percentile is around $500$ and $25$\textsuperscript{th} percentile is around $486$. Similarly, the average \gls{aop} value as shown Fig. \ref{fig:aop_dist_all} and in Table \ref{tab:th_aop} for \gls{vdn} ID$=0$ case is much lower ($5.5$), which is the indication of fairness. Similar conclusion can be drawn for $N=16$ and $N=50$.

The case when IDs $=0$ allows agents update the policies as if it is a single agent (hence single policy). This is unlike the IDs $=1$ case in which, even if the state-space is the same for agents, they behave differently. This way we achieve better trade-off without using agent IDs, which is also scalable and allows devices to join/leave the network without identification. These plots also show that IDs $=1$ case outperforms the \gls{beb} in terms of average throughput but \gls{beb} has better average \gls{aop} than IDs $=1$ case. The average throughput of \gls{beb} technique is higher than IDs $=0$ case for higher values $N$ but no IDs case exhibits much better throughput-fairness tradeoff, as evident from average \gls{aop} values. The case where agents IDs are used is most unfair because of the reward signal that only cares about maximizing the throughput.

Obviously, one can learn a policy by designing a reward function that enforces the devices to be fair even when agent IDs are incorporated; however, such a scenario is not of our interest in this paper.

\subsubsection{Comparison between VDN, QMIX and DRQN}
\begin{figure}
    \centering
    \includegraphics[width=260px, height=180px]{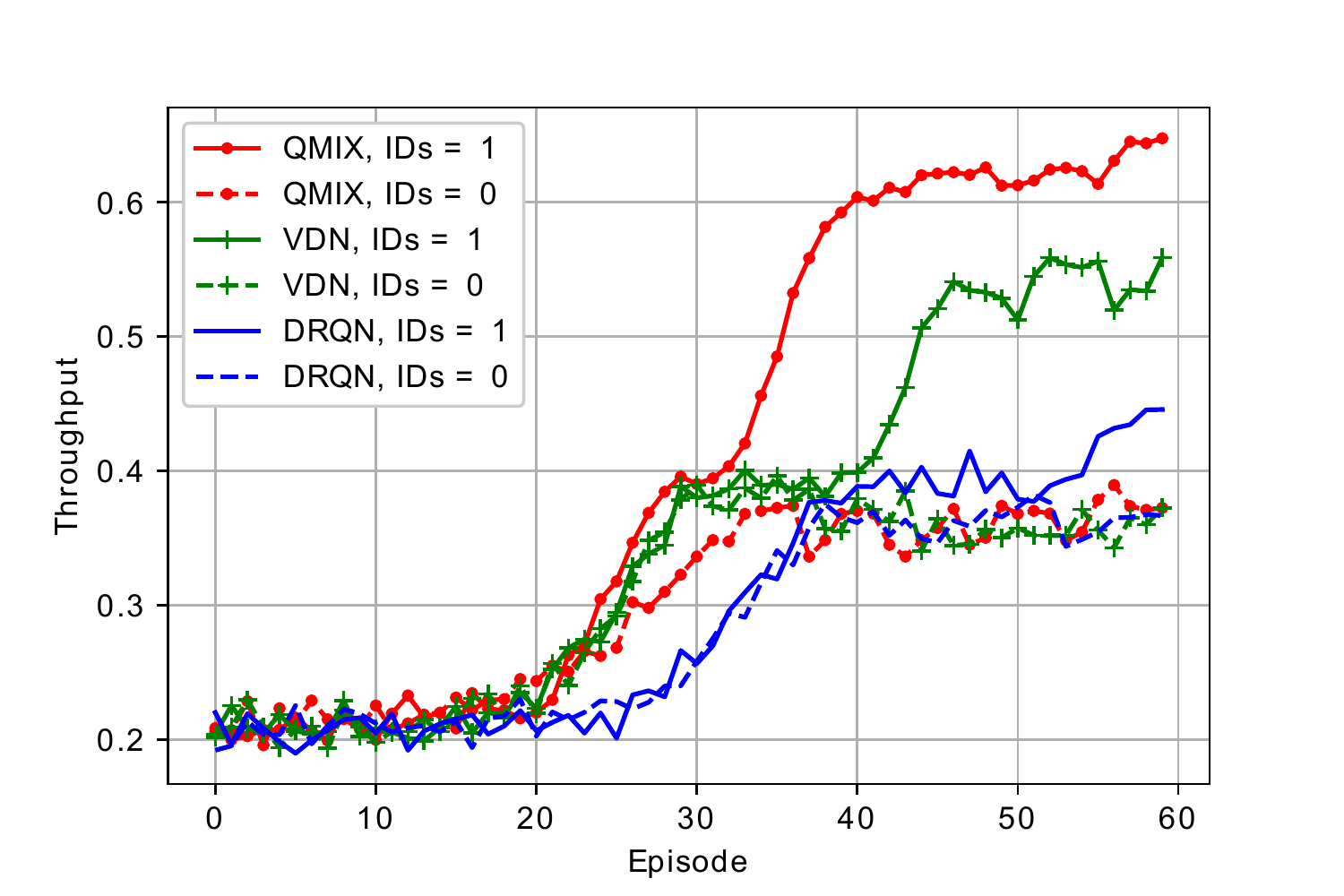}
    \caption{Average throughput comparison of different \gls{marl} algorithms during training, for $N=8$ and $\lambda_n = 0.3$.}
    \label{fig:th_train_marl}
\end{figure}

\begin{figure}
    \centering
    \includegraphics[width=260px, height=220px]{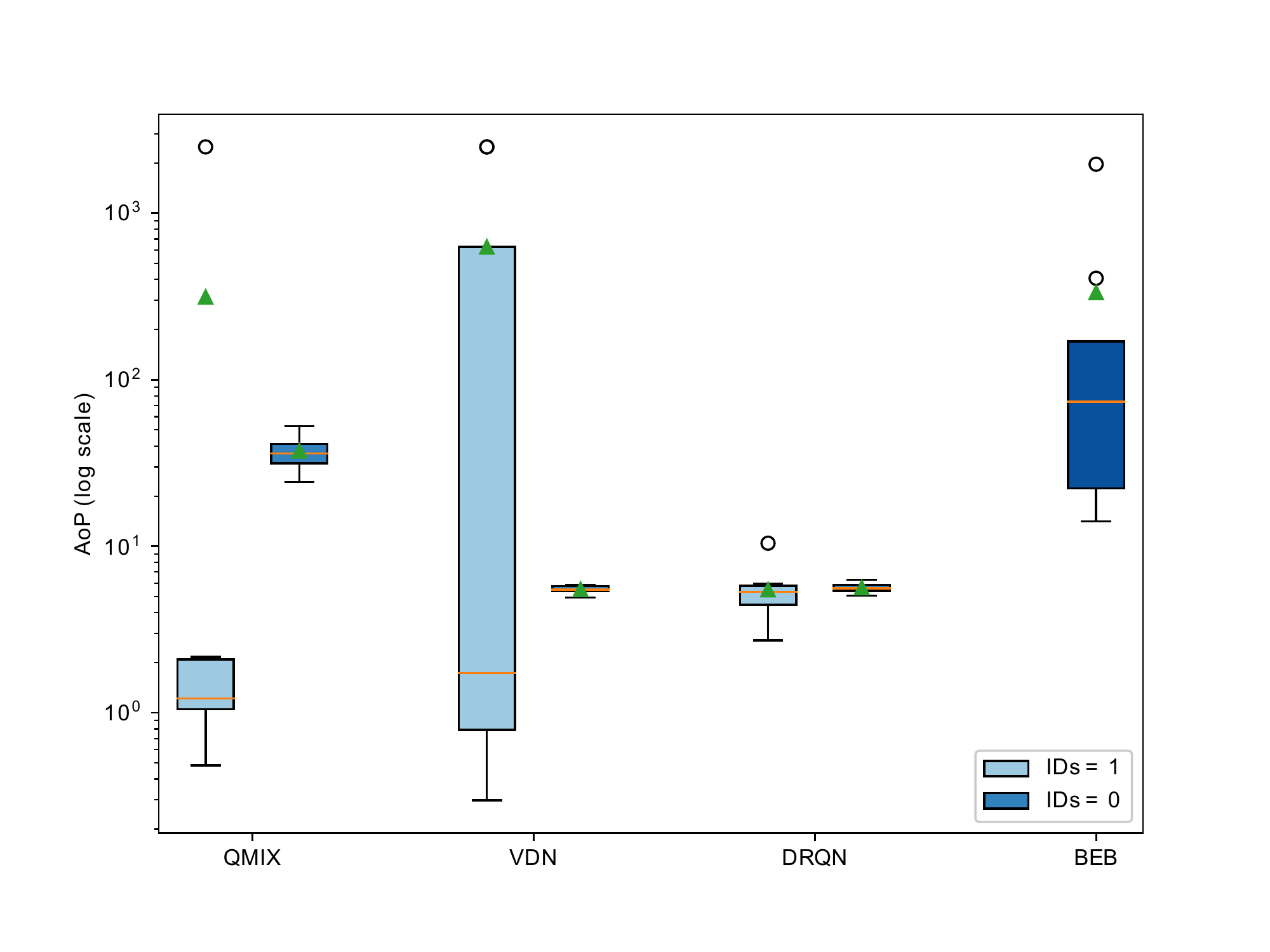}
    \caption{The \gls{aop} comparison among \glspl{mtd} for different \gls{marl} algorithms for $N=8$ and $\lambda_n = 0.3$}
    \label{fig:aop_dist_marl}
\end{figure}

We used \gls{vdn} algorithm to learn \gls{ra} schemes for IDs and no IDs cases. In  Fig. \ref{fig:th_train_marl}, we compare the performance of \gls{vdn} with QMIX and \gls{drqn} for $N=8$ and $M=2$. Both \gls{vdn} and QMIX use mixer networks to calculate the total Q-value $Q_{tot}$ and exploit the benefits of centralized learning. However, \gls{drqn} does not take any such advantage of centralized training. For this reason, we can see that both QMIX and \gls{vdn} algorithms learn a policy that maximizes the throughput for the case when agent IDs are incorporated (IDs $=1$) and QMIX outperforms \gls{vdn}. Interestingly, the \gls{drqn} learns only a slightly better policy when IDs $=1$ as compared to the case when agent IDs $=0$, again, due to the major difference that it doesn't take any advantage of centralization as opposed to the \gls{vdn} and QMIX. However, in Fig. \ref{fig:aop_dist_marl}, we can clearly see that QMIX and \gls{vdn} learn a policy that is unfair when IDs $=1$, \gls{vdn} being more unfair than QMIX. On the other hand, the \gls{drqn} for this case has lower \gls{aop} and relatively much fairer policy than \gls{vdn} and QMIX. It does not imply that \gls{drqn} is a better algorithm than \gls{vdn} and QMIX. In fact, QMIX outperforms \gls{vdn} and both of them outperform \gls{drqn} as far as the objective (maximizing throughput) is concerned.  Another interesting observation is that the learned policies are very similar between all the algorithms for IDs $=0$ case and it is evident from both Fig. \ref{fig:th_train_marl} and Fig. \ref{fig:aop_dist_marl}. They are fair but it seems that exploiting centralization advantages without using agent IDs does not provide significant improvements. 

\begin{figure}
    \centering
    \includegraphics[width=180px, height=130px]{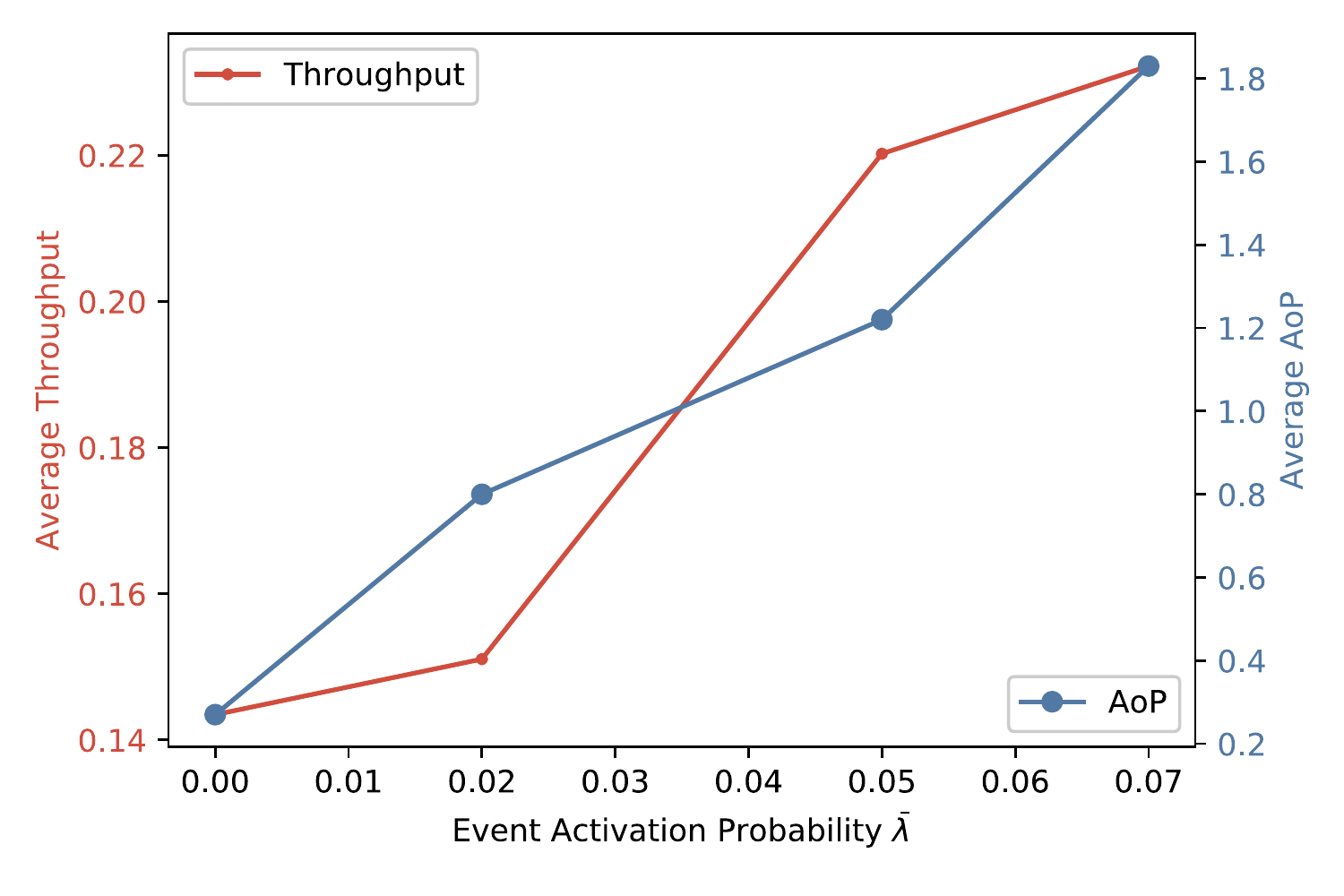}
    \caption{Average throughput and average AoP for different values of $\bar{\lambda}$ when $N=20$, $L=3$ and $\lambda = 0.3$}
    \label{fig:ed_aop_th}
\end{figure}

\subsection{RESULTS FOR CORRELATED TRAFFIC}
In this work, we are interested in designing an access policy for the \glspl{mtd} deployed in an area, and whenever an event $l$ happens, the \glspl{mtd} in the vicinity of the event or the \glspl{mtd} closer to the epicenter of the event become active in a correlated manner. To model this, we calculate the probability of a device at location $\mathbf{x}$ becoming active due to the event happening at epicenter $\mathbf{y}$ as $p_{\mathbf{x}\mathbf{y}}$ in the following way:
\begin{align}
    p_{\mathbf{x}\mathbf{y}} = 
    \begin{cases}
        1 \quad \text{if } d_{\mathbf{x}\mathbf{y}} \leq d_{th} \\
        0 \quad \text{otherwise},
    \end{cases}
\end{align}
where $d_{\mathbf{x}\mathbf{y}} = \lVert  \mathbf{x}-\mathbf{y}  \rVert$ and $d_{th}$ is the threshold distance. We assume that the events are atomic in nature, i.e., if an event becomes active in time slot $k$, it activates \glspl{mtd} within $d_{th}$. We assume that each \gls{mtd} activated by an event has one packet each to transmit and the \glspl{mtd} remain active until their packets are successfully transmitted. 
\begin{figure*}
\centering
        \begin{subfigure}{.45\textwidth}
        \includegraphics[width=240px, height=180px]{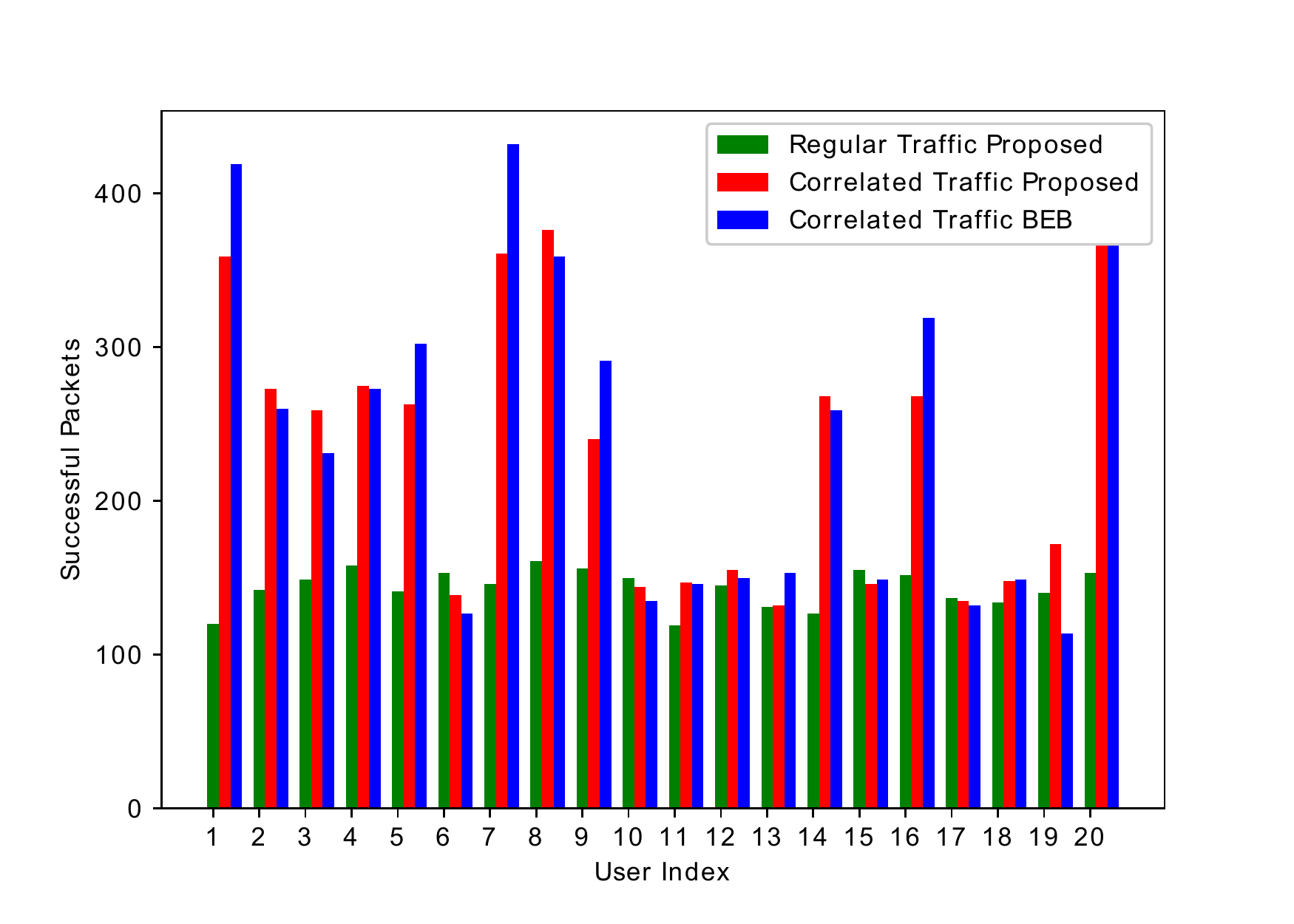}
        \caption{Successful packets per user}
        \label{fig:ed_pkt_dist_8u}
        \end{subfigure}
        \begin{subfigure}{.45\textwidth}
        \includegraphics[width=240px, height=180px]{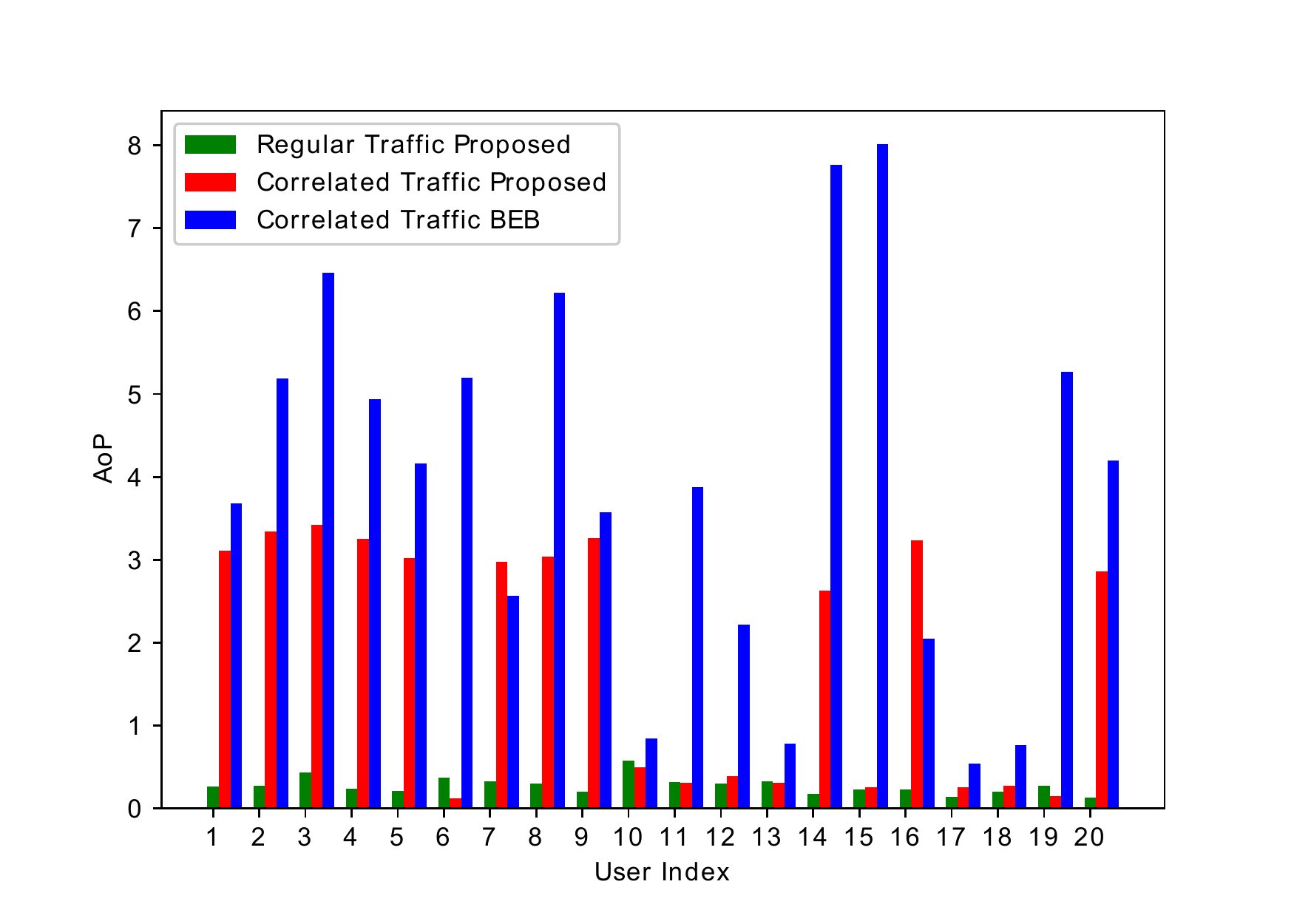}
        \caption{AoP per user}
        \label{fig:ed_aop_dist_8u}
        \end{subfigure}
    \caption{The comparison of successful packets per user and AoP of each user for \gls{ed} traffic model for $(K, N)= (10000, 20)$, $L=3$, $\lambda = 0.3$ and $\bar{\lambda} = 0.07$.}
    \label{fig:ed_pkt_aop_dist_8u}
\end{figure*}

For correlated traffic arrival, we consider the example as shown in Fig. \ref{fig:ed_reg_traffic}, in which $N=20$ \glspl{mtd} are randomly distributed in rectangular area. The \glspl{mtd} follow Bernoulli process with average arrival rate $\lambda = 0.3$ for regular traffic. We consider $L=3$ event epicenters and \glspl{mtd} belonging to each epicenter are given in \ref{tab:event_usr_index}. The events become active following another independent Bernoulli process with average event activation probability given by $\bar{\lambda}$ such that $p = \bar{\lambda}/L$. We train and test for different values of $\bar{\lambda}$ as shown in Table \ref{tab:event_activ_exp}. The training for each $\bar{\lambda}$ is performed over $60$ episodes and $K=2000$ time slots per episode. 

Since we want to learn the same policy for each agent, we do not use agent IDs for correlated traffic arrivals case and we only consider \gls{vdn} IDs $=0$ case since have seen that \gls{vdn} performs better as compared to the QMIX and \gls{drqn}, Moreover, we do not consider whether events-driven traffic has any priority over regular traffic. All events are of the same nature and the same reward function as the regular traffic is used. Fig. \ref{fig:ed_aop_th} shows the average throughput and average \gls{aop} for different values of $\bar{\lambda}$. The value $\bar{\lambda} = 0$ means that there is only regular traffic and we see in Fig. \ref{fig:ed_aop_th} that as the $\bar{\lambda}$ increases, the average throughput and average AoP both increase, which is natural because when there is more traffic, there are more packets being successful but require more time to be transmitted. 
\begin{table}
\centering
    \caption{Number of times each event was activated for $K = 10,000$ time slots for different event activation probabilities $\bar{\lambda}$.}
    \renewcommand{\arraystretch}{1.0}
    \begin{tabular}{|l|l|l|l|}
    \hline
    $\bar{\lambda}$  & \textbf{Event 1} & \textbf{Event 2} & \textbf{Event 3}   
     \\ \hline
    $0.02$       & 63 & 61 &59      
    \\ \hline
    $0.05$ &170 &170 &169
    \\ \hline
    $0.07$ &243 &226 &241
    \\ \hline
    \end{tabular}

\label{tab:event_activ_exp}
\end{table}
\begin{table}
\centering
\caption{\glspl{mtd} reporting the events as in Fig. \ref{fig:ed_reg_traffic}}
\renewcommand{\arraystretch}{1.0}
    \begin{tabular}{|l|l|}
    \hline
    Event \#  & \glspl{mtd} Index 
     \\ \hline
    $1$       & $1,7,8,16,20$      
    \\ \hline
    $2$ & $2,3,4,5,9, 14$
    \\ \hline
    $3$ & $1,7,8,20$
    \\ \hline
    \end{tabular}

\label{tab:event_usr_index}
\end{table}
By further zooming in on individual \glspl{mtd}, we want to observe how each \gls{mtd} is behaving in correlated traffic scenario. In this case, we do not incorporate agent IDs in the state-space of agents. Fig. \ref{fig:ed_pkt_aop_dist_8u} shows successful number of packets and \gls{aop} per device and we compare the correlated traffic with regular traffic arrival scenario. Clearly, only users that are involved in reporting any event have higher throughput as well as higher \gls{aop}. Obviously, when few users become active together, they will take more time to resolve collision and to send their packets successfully. The reason to show these plots is that the \gls{rl}-based algorithms adapt to the traffic changes as the devices that are not involved in reporting any events have similar \gls{aop} and packet success rate to the regular traffic case, and only the \glspl{mtd} belonging to events change their policies. The baseline \gls{beb} does not really adapt or care whether the devices are involved in events. The throughput is high for \gls{beb} for devices that are receiving more packets which is not surprising but if we look at the \gls{aop} plot in Fig. \ref{fig:ed_aop_dist_8u}, there are devices that are not involved in reporting any events but they have higher \gls{aop} for \gls{beb} as opposed to the proposed algorithm. 

\subsection{SCALABILITY AND ROBUSTNESS ANALYSIS}
\begin{figure}
    \centering
    \includegraphics[width=240px, height=170px]{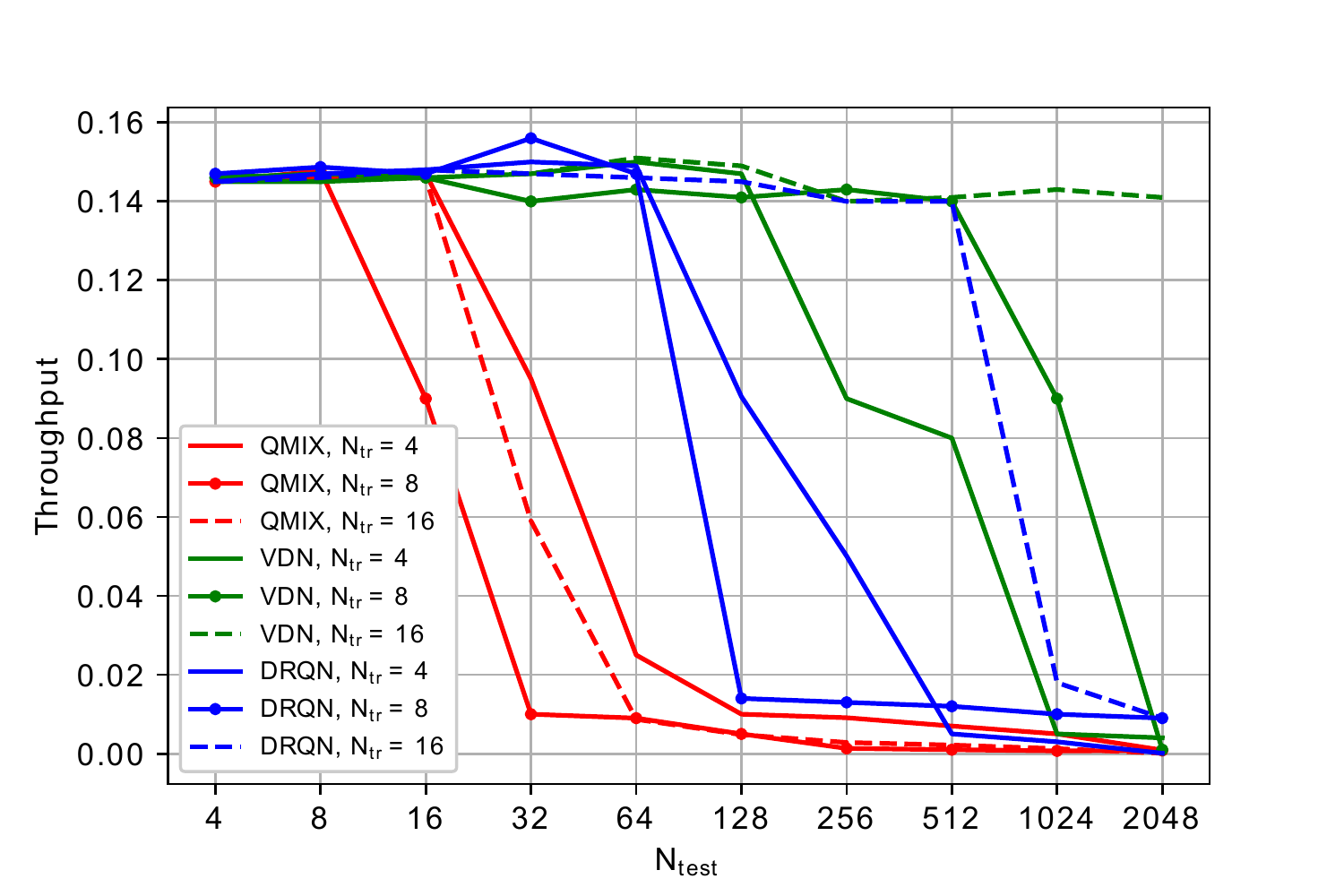}
    \caption{Average Throughput performance of the learned policy for $\lambda = 0.3$ for different $N_{\mathrm{tr}}$ and tested for $N_{\mathrm{test}}$.}
    \label{fig:scaled_N_th}
\end{figure}

We compare the learned policies of \gls{vdn}, QMIX and \gls{drqn} for no IDs case and show how robust the policy learned is by each algorithm if we scale it for a higher number of devices. The performance of each algorithm in terms of average throughput is shown in Fig. \ref{fig:scaled_N_th}. We denote with $N_{\mathrm{tr}}$ the number of devices during training, and the number of devices for testing is denoted by $N_{\mathrm{test}}$. The average arrival rate of the system is $\lambda = 0.3$. The results are simulated for $3$ different random seeds and the best performances are shown in Fig. \ref{fig:scaled_N_th}.

We show that the \gls{vdn} has more robust policy than both QMIX and \gls{drqn}. For $\lambda = 0.3$, the policy learned for $N_{\mathrm{tr}}=4$ performs the same $N_{\mathrm{test}} = {4, 8, 16}$ and the throughput starts dropping after that. It is because the policy learned for $N_{\mathrm{tr}}=4$ has higher $\lambda_n = \lambda/N_{\mathrm{tr}}$ and as the number of devices grow, the collisions are not resolved and hence the average throughput drops to almost $0$. On the other hand, the policy learned for a relatively higher number of devices such as $N_{\mathrm{tr}} = 16$ is robust for the number of devices less than $N_{\mathrm{tr}}$ and also scales for a large number of devices. 

Intuitively, for instance, when $N_{\mathrm{tr}}=4$ and $\lambda=0.3$, then $\lambda_n$ for smaller $N_{\mathrm{tr}}$ has higher arrival rate or in other words, the \glspl{mtd} observe packet arrival more frequently than $\lambda_n$ for larger $N_{\mathrm{tr}}$ and therefore, devices learn to be more aggressive in terms of their transmissions to empty their buffers and such policy does not perform well for a very large number of devices.

Furthermore, the policy of QMIX has worse performance as compared to both \gls{vdn} and \gls{drqn}. We can also observe that QMIX without incorporating agent IDs is not as effective and not as robust as compared to the \gls{vdn}. 
\section{CONCLUSION AND FUTURE WORK}
In this paper, we have proposed \gls{marl}-based \gls{ra} schemes for multi-user multi-channel \gls{mmtc} network. We have used the broadcast feedback commonly received by all the \glspl{mtd}. We demonstrated that incorporating agents IDs is not suitable for \gls{mmtc} systems, as we aim to design a fair and a scalable scheme. We have shown that even when the optimization objective is to maximize throughput, not incorporating agent IDs provides a fair use of resources by each agent and, thus, results in a better throughput-fairness tradeoff as compared to the \gls{beb} scheme and when IDs are used. This is supported by our analysis of successful packet distribution per user 
and also through average \gls{aop}. We presented the scalability analysis of the proposed algorithms where we omit agent IDs for learning and we conclude that our system scales well for lower average arrival rates and that the learned policy is more robust for \gls{vdn} as compared to the QMIX and \gls{drqn}. Moreover, we have presented the suitability of several popular \gls{marl} algorithms and have shown that the \gls{vdn} and QMIX take advantage of centralization and they are better suited for designing \gls{ra} schemes for \gls{mmtc}. We have demonstrated that \gls{rl}-based algorithms can adapt to changes in traffic, whereas \gls{eb} schemes are not aware of such changes. We have used a correlated traffic arrival model along with the regular traffic, for which we show that the users learn the correlation and adapt to  changes in the traffic. 

For the correlated traffic, we have assumed that all the events are of the same nature and they have the same priority. Future works could consider prioritizing events to learn a scheme where the devices with high priority send their packets with low latency. Moreover, one can use agent IDs for a system where the devices are fixed and known, we can design a reward that is fair and that exploits correlation between devices even better. Furthermore, whilst omitting IDs make the scheme fair, the system throughput goes to zero as for a large number of devices; therefore, there may be a need to design algorithms with some coordination among devices or a group of devices. 


\section*{ACKNOWLEDGMENT}
This work was partially supported by the European Union H2020 Research and Innovation Programme through Marie Skłodowska Curie action (MSCA-ITN-ETN 813999 WINDMILL), and Spanish grant PID 2021-128373OB-I00. 

\bibliographystyle{ieeetr}
\bibliography{bibliography}

\begin{thebibliography}{10}

\bibitem{five_5g}
F.~Boccardi, R.~W. Heath, A.~Lozano, T.~L. Marzetta, and P.~Popovski, ``Five
  disruptive technology directions for {5G},'' {\em IEEE Communications
  Magazine}, vol.~52, no.~2, pp.~74--80, 2014.

\bibitem{bockelmann_18}
C.~Bockelmann, N.~K. Pratas, G.~Wunder, S.~Saur, M.~Navarro, D.~Gregoratti,
  G.~Vivier, E.~De~Carvalho, Y.~Ji, C.~Stefanovic, P.~Popovski, Q.~Wang,
  M.~Schellmann, E.~Kosmatos, P.~Demestichas, M.~Raceala-Motoc, P.~Jung,
  S.~Stanczak, and A.~Dekorsy, ``Towards massive connectivity support for
  scalable {mMTC} communications in 5{G} networks,'' {\em IEEE Access}, vol.~6,
  pp.~28969--28992, 2018.

\bibitem{bockelmann_16}
C.~Bockelmann, N.~Pratas, H.~Nikopour, K.~Au, T.~Svensson, C.~Stefanovic,
  P.~Popovski, and A.~Dekorsy, ``Massive machine-type communications in 5{g}:
  physical and {MAC}-layer solutions,'' {\em IEEE Communications Magazine},
  vol.~54, no.~9, pp.~59--65, 2016.

\bibitem{nbiot_ltem}
``Standards for the {IoT}.''
  \url{https://www.3gpp.org/news-events/1805-iot_r14}.
\newblock Accessed: 2022-09-20.

\bibitem{sigfox}
\url{https://www.sigfox.com/}.

\bibitem{lora}
\url{https://lora-alliance.org/}.

\bibitem{ra_mtc_israel_19}
I.~Leyva-Mayorga, C.~Stefanovic, P.~Popovski, V.~Pla, and J.~Martinez-Bauset,
  {\em Random Access for Machine-Type Communications}.
\newblock United Kingdom: Wiley, Dec. 2019.

\bibitem{beb_analysis}
B.-J. Kwak, N.-O. Song, and L.~Miller, ``Performance analysis of exponential
  backoff,'' {\em IEEE/ACM Transactions on Networking}, vol.~13, no.~2,
  pp.~343--355, 2005.

\bibitem{Barletta18}
L.~Barletta, F.~Borgonovo, and I.~Filippini, ``The throughput and access delay
  of slotted-aloha with exponential backoff,'' {\em IEEE/ACM Transactions on
  Networking}, vol.~26, no.~1, pp.~451--464, 2018.

\bibitem{Naparstek}
O.~{Naparstek} and K.~{Cohen}, ``Deep multi-user reinforcement learning for
  distributed dynamic spectrum access,'' {\em IEEE Transactions on Wireless
  Communications}, vol.~18, no.~1, pp.~310--323, 2019.

\bibitem{wang2018}
S.~Wang, H.~Liu, P.~H. Gomes, and B.~Krishnamachari, ``Deep reinforcement
  learning for dynamic multichannel access in wireless networks,'' {\em IEEE
  Transactions on Cognitive Communications and Networking}, vol.~4, no.~2,
  pp.~257--265, 2018.

\bibitem{HLi}
H.~{Li}, ``Multi-agent {Q}-learning for competitive spectrum access in
  cognitive radio systems,'' in {\em 2010 Fifth IEEE Workshop on Networking
  Technologies for Software Defined Radio Networks (SDR)}, pp.~1--6, 2010.

\bibitem{challita_drl}
U.~Challita, L.~Dong, and W.~Saad, ``Proactive resource management for {LTE} in
  unlicensed spectrum: A deep learning perspective,'' {\em IEEE Transactions on
  Wireless Communications}, vol.~17, no.~7, pp.~4674--4689, 2018.

\bibitem{CHU201523}
Y.~Chu, S.~Kosunalp, P.~D. Mitchell, D.~Grace, and T.~Clarke, ``Application of
  reinforcement learning to medium access control for wireless sensor
  networks,'' {\em Engineering Applications of Artificial Intelligence},
  vol.~46, pp.~23--32, 2015.

\bibitem{alfaro_alohaq}
L.~{de Alfaro}, M.~{Zhang}, and J.~J. {Garcia-Luna-Aceves}, ``Approaching fair
  collision-free channel access with slotted aloha using collaborative
  policy-based reinforcement learning,'' in {\em 2020 IFIP Networking
  Conference (Networking)}, pp.~262--270, 2020.

\bibitem{zhong2019deep}
C.~Zhong, Z.~Lu, M.~C. Gursoy, and S.~Velipasalar, ``Actor-critic deep
  reinforcement learning for dynamic multichannel access,'' in {\em 2018 IEEE
  Global Conference on Signal and Information Processing (GlobalSIP)},
  pp.~599--603, 2018.

\bibitem{xu_milcom}
Y.~Xu, J.~Yu, W.~Headley, and R.~Buehrer, ``Deep reinforcement learning for
  dynamic spectrum access in wireless networks,'' in {\em MILCOM 2018 - 2018
  IEEE Military Communications Conference (MILCOM)}, pp.~207--212, 2018.

\bibitem{tomovic_20}
S.~Tomovic and I.~Radusinovic, ``A novel deep {Q}-learning method for dynamic
  spectrum access,'' in {\em 2020 28th Telecommunications Forum (TELFOR)},
  pp.~1--4, 2020.

\bibitem{rapid}
J.~Kim, S.~Kim, T.~Taleb, and S.~Choi, ``{RAPID}: Contention resolution based
  random access using context {ID} for {IoT},'' {\em IEEE Transactions on
  Vehicular Technology}, vol.~68, no.~7, pp.~7121--7135, 2019.

\bibitem{nan_21}
N.~Jiang, Y.~Deng, A.~Nallanathan, and J.~Yuan, ``A decoupled learning strategy
  for massive access optimization in cellular {IoT} networks,'' {\em IEEE
  Journal on Selected Areas in Communications}, vol.~39, no.~3, pp.~668--685,
  2021.

\bibitem{zhang_doubleRA22}
C.~Zhang, X.~Sun, W.~Xia, J.~Zhang, H.~Zhu, and X.~Wang, ``Deep learning based
  double-contention random access for massive machine-type communications,''
  {\em IEEE Transactions on Wireless Communications}, pp.~1--1, 2022.

\bibitem{alberto_21}
A.~Rech and S.~Tomasin, ``Coordinated random access for industrial iot with
  correlated traffic by reinforcement-learning,'' in {\em 2021 IEEE Globecom
  Workshops (GC Wkshps)}, pp.~1--6, 2021.

\bibitem{jadoon_wcnc22}
M.~A. Jadoon, A.~Pastore, M.~Navarro, and F.~Perez-Cruz, ``Deep reinforcement
  learning for random access in machine-type communication,'' in {\em 2022 IEEE
  Wireless Communications and Networking Conference (WCNC)}, pp.~2553--2558,
  2022.

\bibitem{jadoon_vtc}
M.~A. Jadoon, A.~Pastore, and M.~Navarro, ``Collision resolution with deep
  reinforcement learning for random access in machine-type communication,'' in
  {\em 2022 IEEE 95th Vehicular Technology Conference: (VTC2022-Spring)},
  pp.~1--6, 2022.

\bibitem{8Yu2019het}
Y.~Yu, T.~Wang, and S.~C. Liew, ``Deep-reinforcement learning multiple access
  for heterogeneous wireless networks,'' {\em IEEE Journal on Selected Areas in
  Communications}, vol.~37, no.~6, pp.~1277--1290, 2019.

\bibitem{acb_nbiot}
Y.~Hadjadj-Aoul and S.~Ait-Chellouche, ``Access control in {NB-IoT} networks: A
  deep reinforcement learning strategy,'' {\em Information}, vol.~11, no.~11,
  2020.

\bibitem{Shao_22}
Y.~Shao, Y.~Cai, T.~Wang, Z.~Guo, P.~Liu, J.~Luo, and D.~Gunduz,
  ``Learning-based autonomous channel access in the presence of hidden
  terminals,'' 2022.

\bibitem{Anders_18}
A.~E. Kalor, O.~A. Hanna, and P.~Popovski, ``Random access schemes in wireless
  systems with correlated user activity,'' in {\em 2018 IEEE 19th International
  Workshop on Signal Processing Advances in Wireless Communications (SPAWC)},
  pp.~1--5, 2018.

\bibitem{zheng_22}
C.~Zheng, M.~Egan, L.~Clavier, A.~E. Kalør, and P.~Popovski, ``Stochastic
  resource allocation for outage minimization in random access with correlated
  activation,'' in {\em 2022 IEEE Wireless Communications and Networking
  Conference (WCNC)}, pp.~1635--1640, 2022.

\bibitem{federico_21}
F.~Moretto, A.~Brighente, and S.~Tomasin, ``Greedy maximum- throughput
  grant-free random access for correlated {IoT} traffic,'' in {\em 2021 IEEE
  94th Vehicular Technology Conference (VTC2021-Fall)}, pp.~1--5, 2021.

\bibitem{Thomsen17}
H.~Thomsen, C.~N. Manchon, and B.~H. Fleury, ``A traffic model for machine-type
  communications using spatial point processes,'' in {\em 2017 IEEE 28th Annual
  International Symposium on Personal, Indoor, and Mobile Radio Communications
  (PIMRC)}, pp.~1--6, 2017.

\bibitem{mnih2015humanlevel}
V.~Mnih {\em et~al.}, ``Human-level control through deep reinforcement
  learning,'' {\em Nature}, vol.~518, pp.~529--533, Feb. 2015.

\bibitem{maddpg_LoweWTHAM17}
R.~Lowe, Y.~Wu, A.~Tamar, J.~Harb, P.~Abbeel, and I.~Mordatch, ``Multi-agent
  actor-critic for mixed cooperative-competitive environments,'' {\em CoRR},
  vol.~abs/1706.02275, 2017.

\bibitem{mappo_chao2021}
C.~Yu, A.~Velu, E.~Vinitsky, Y.~Wang, A.~M. Bayen, and Y.~Wu, ``The surprising
  effectiveness of {MAPPO} in cooperative, multi-agent games,'' {\em CoRR},
  vol.~abs/2103.01955, 2021.

\bibitem{qmix}
T.~Rashid, M.~Samvelyan, C.~S. De~Witt, G.~Farquhar, J.~Foerster, and
  S.~Whiteson, ``Monotonic value function factorisation for deep multi-agent
  reinforcement learning,'' {\em J. Mach. Learn. Res.}, vol.~21, jun 2022.

\bibitem{coma}
J.~N. Foerster, G.~Farquhar, T.~Afouras, N.~Nardelli, and S.~Whiteson,
  ``Counterfactual multi-agent policy gradients,'' AAAI'18/IAAI'18/EAAI'18,
  AAAI Press, 2018.

\bibitem{vdn}
P.~Sunehag, G.~Lever, A.~Gruslys, W.~M. Czarnecki, V.~Zambaldi, M.~Jaderberg,
  M.~Lanctot, N.~Sonnerat, J.~Z. Leibo, K.~Tuyls, and T.~Graepel,
  ``Value-decomposition networks for cooperative multi-agent learning based on
  team reward,'' in {\em Proceedings of the 17th International Conference on
  Autonomous Agents and MultiAgent Systems}, AAMAS '18, (Richland, SC),
  p.~2085–2087, International Foundation for Autonomous Agents and Multiagent
  Systems, 2018.

\bibitem{mf_marl}
Y.~Yang, R.~Luo, M.~Li, M.~Zhou, W.~Zhang, and J.~Wang, ``Mean field
  multi-agent reinforcement learning,'' in {\em Proceedings of the 35th
  International Conference on Machine Learning} (J.~Dy and A.~Krause, eds.),
  vol.~80 of {\em Proceedings of Machine Learning Research}, pp.~5571--5580,
  PMLR, 10--15 Jul 2018.

\bibitem{maac_iqbal}
S.~Iqbal and F.~Sha, ``Actor-attention-critic for multi-agent reinforcement
  learning,'' 2018.

\bibitem{soft_ac}
T.~Haarnoja, A.~Zhou, P.~Abbeel, and S.~Levine, ``Soft actor-critic: Off-policy
  maximum entropy deep reinforcement learning with a stochastic actor,'' 2018.

\bibitem{mateus_vtc22}
M.~P. Mota, A.~Valcarce, and J.-M. Gorce, ``Scalable joint learning of wireless
  multiple-access policies and their signaling,'' in {\em 2022 IEEE 95th
  Vehicular Technology Conference: (VTC2022-Spring)}, pp.~1--5, 2022.

\bibitem{Vouros22}
T.~Kravaris and G.~A. Vouros, ``Deep multiagent reinforcement learning methods
  addressing the scalability challenge,'' in {\em Multi-Agent Technologies and
  Machine Learning} (D.~I. Sheremet, ed.), ch.~21, Rijeka: IntechOpen, 2022.

\bibitem{gupta_2017}
J.~K. Gupta, M.~Egorov, and M.~Kochenderfer, ``Cooperative multi-agent control
  using deep reinforcement learning,'' in {\em Autonomous Agents and Multiagent
  Systems} (G.~Sukthankar and J.~A. Rodriguez-Aguilar, eds.), (Cham),
  pp.~66--83, Springer International Publishing, 2017.

\bibitem{qtran_19}
K.~Son, D.~Kim, W.~J. Kang, D.~E. Hostallero, and Y.~Yi, ``{QTRAN}: Learning to
  factorize with transformation for cooperative multi-agent reinforcement
  learning,'' in {\em Proceedings of the 36th International Conference on
  Machine Learning} (K.~Chaudhuri and R.~Salakhutdinov, eds.), vol.~97 of {\em
  Proceedings of Machine Learning Research}, pp.~5887--5896, PMLR, 09--15 Jun
  2019.

\end{thebibliography}

\newpage

\begin{IEEEbiography}[{\includegraphics[width=1in,height=1.25in,clip,keepaspectratio]{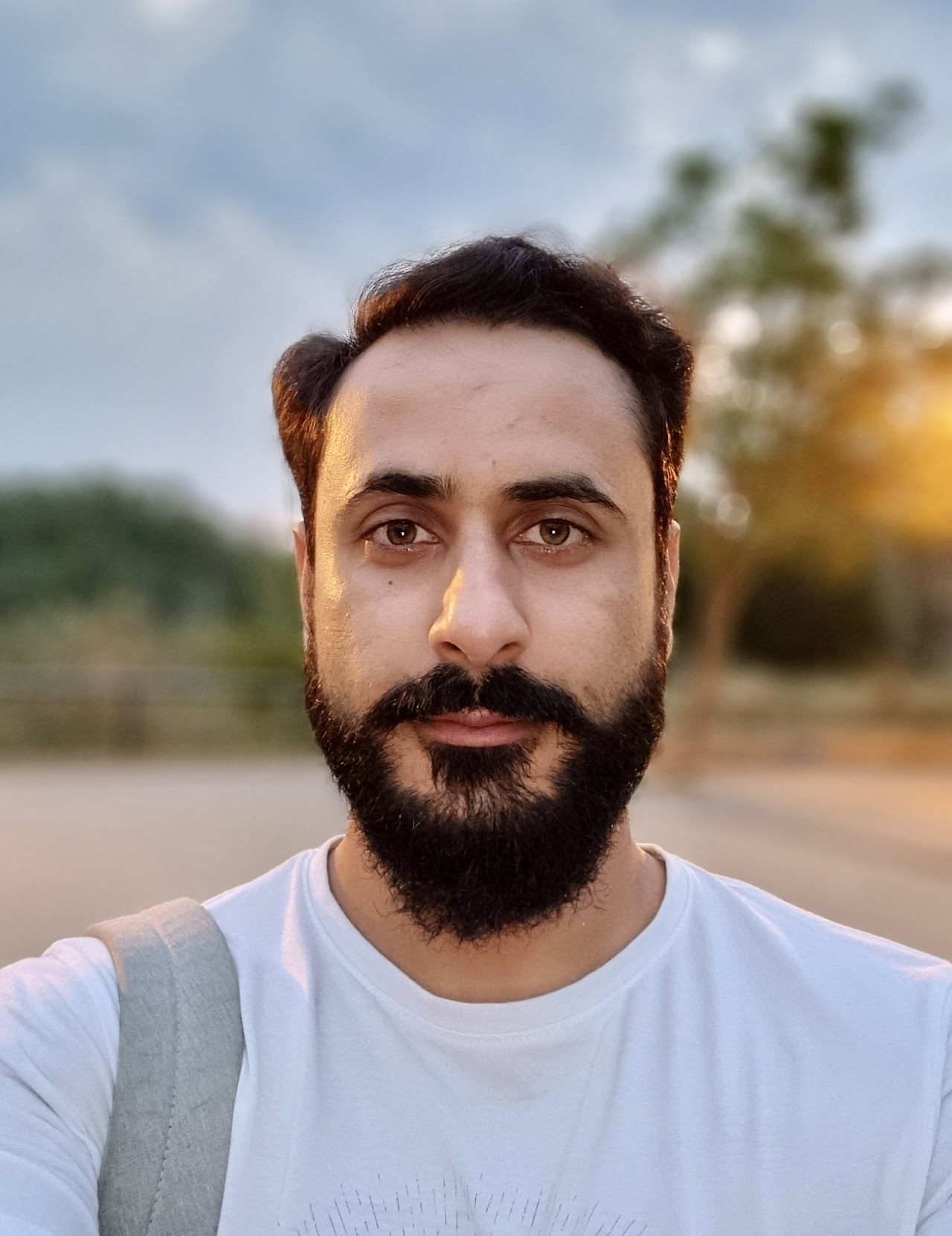}}]
{MUHAMMAD A. JADOON}~(Student Member, IEEE)~is a research assistant at the Centre Tecnològic de Telecomunicacions de Catalunya (CTTC) and an early stage researcher (ESR) of ITN Windmill. He is also a PhD student at the department of Signal Theory and Communications (TSC) at the Universitat Politècnica de Catalunya (UPC). He received his MSc degree in electrical engineering from the University of Ulsan South Korea and his BSc degree in telecommunication engineering from the University of Engineering and Technology Peshawar Pakistan.  

Muhammad's current research is at the intersection of wireless communication and machine learning, with a focus on multi-agent reinforcement learning for resource management in massive machine-type communication. As an ESR of ITN Windmill, he has had the opportunity to work at the Swiss Data Science Center at ETH Zurich and Nokia Bell Labs Paris for his academic and industrial secondments, respectively.\end{IEEEbiography}

\begin{IEEEbiography}[{\includegraphics[width=1in,height=1.25in,clip,keepaspectratio]{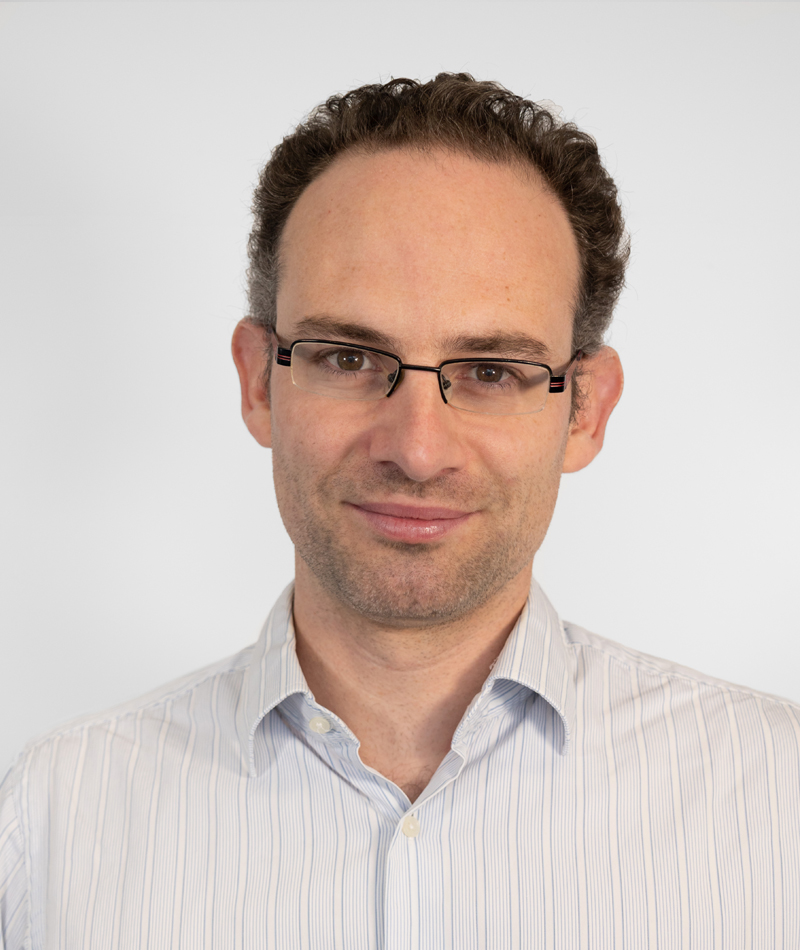}}]
{A. PASTORE,}~(Senior Member, IEEE)~is a Senior Researcher at the Centre Tecnològic de Telecomunicacions de Catalunya, within the Research Unit on Information and Signal Processing for Intelligent Communications. He received a Diplôme de l’École Centrale Paris (ECP, now CentraleSupélec) in 2006 and a Dipl.-Ing.\ degree in electrical engineering in 2009 from the Technical University of Munich, and obtained his PhD from the Universitat Politècnica de Catalunya in 2014. From 2014 to 2016 he has been a postdoctoral researcher at École Polytechnique Fédérale de Lausanne (EPFL) in the Laboratory for Information in Networked Systems (LINX) headed by Prof. Michael Gastpar.

His topics of interest lie mainly in the fields of information theory and signal processing for wireless communications, machine learning for communications, physical-layer network coding, protocol learning, quantum key distribution, and privacy--utility tradeoffs.
\end{IEEEbiography}

\begin{IEEEbiography}[{\includegraphics[width=1in,height=1.25in,clip,keepaspectratio]{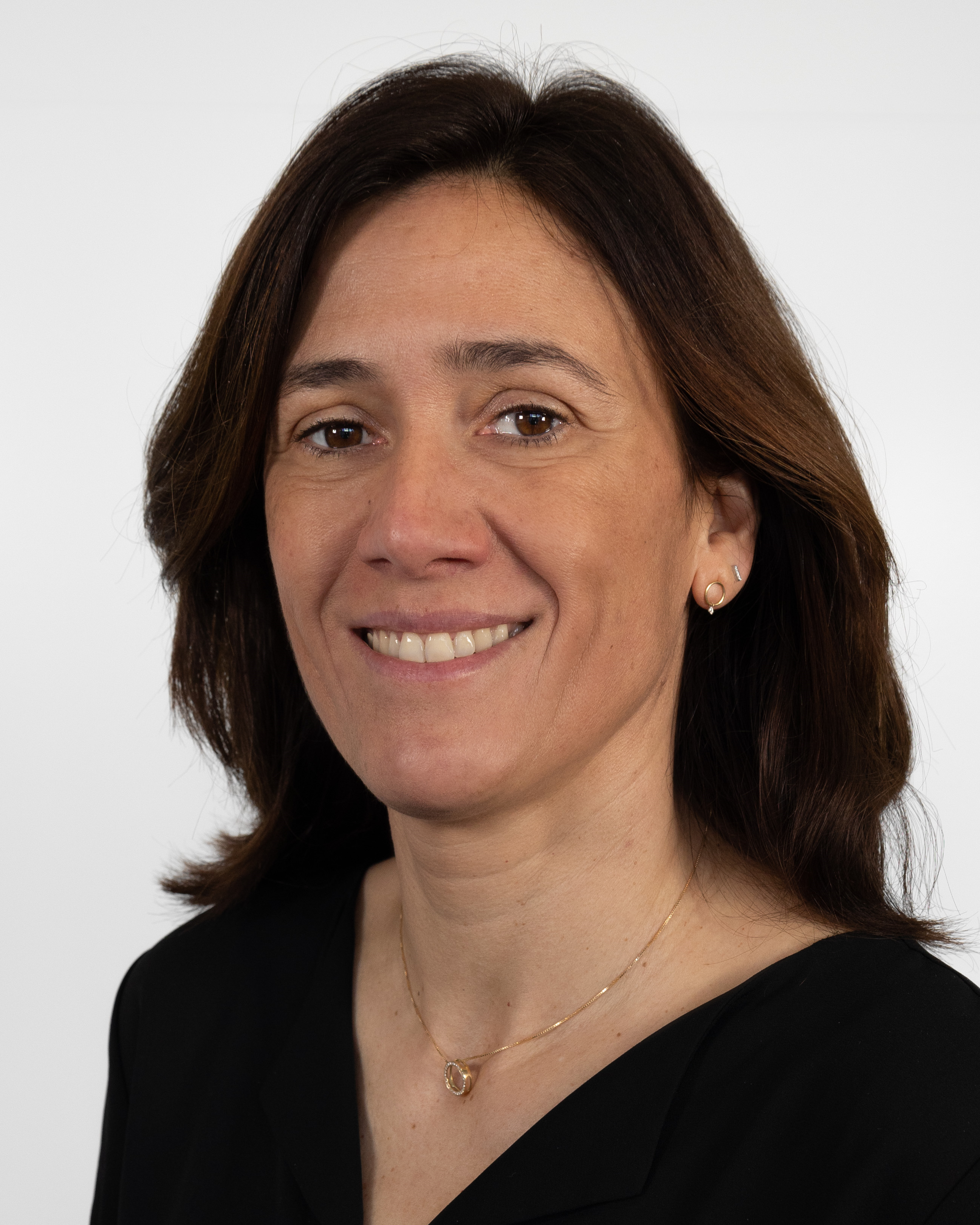}}]
{M. NAVARRO,}~(Senior Member, IEEE)~is a Senior Researcher at the Centre Tecnològic de Telecomunicacions de Catalunya, where she led the Communication Systems Division group from 2016 to 2021 and is currently part of the Direction unit. She received the MSc degree in Telecommunications Engineering from Universitat Politècnica de Catalunya in 1997 and the PhD degree in Telecommunications from the Institute for Telecommunications Research (ITR), University of South Australia, in 2002. From Oct. 1997 to Dec. 1998 she was a Research Assistant at the Department of Signal Theory and Communications at the UPC, where she worked on the development of fractal shape multiband antennas for wireless cellular communications systems. She has also been part-time lecturer at the Universitat Pompeu Fabra, Barcelona. Over the last 15 years she has lead projects funded by the European Commission, Spanish and Catalan Governments, as well as the European Space Agency (ESA), spanning across 3G to 5G air interface designs, modem prototypes for space applications, virtualized wireless networks or intelligent transport systems. Her primary areas of interest are on digital communications and information processing with applications to wireless communications and positioning. She served at the Editorial Board of Emerging Telecommunications Technologies (ETT). 
\end{IEEEbiography}

\begin{IEEEbiography}[{\includegraphics[width=1in,height=1.25in,clip,keepaspectratio]{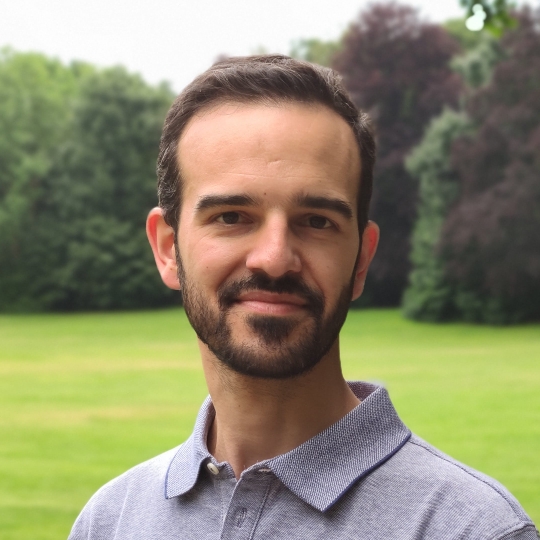}}]
{A. VALCARCE,}~(Senior Member, IEEE)~is Head of Department on Wireless AI/ML at Nokia Bell Labs, France. His research is focused on the application of machine learning techniques to L2 and L3 wireless problems for the development of technologies beyond 5G. He is especially interested on the potential of multiagent reinforcement learning for emerging novel L2 signaling protocols, as well as on the usage of Bayesian optimization for RRM problems. His background is on cellular networks, computational electromagnetics, optimization algorithms, and machine learning.
\end{IEEEbiography}

\vfill\pagebreak

\end{document}